\documentclass[twocolumn]{aastex631}
\pdfoutput=1

\usepackage{amsmath} 
\usepackage{natbib}
\usepackage{subfigure}
\usepackage{graphicx}  
\usepackage{float} 
\usepackage{threeparttable}

\begin{document}

\title{Searching for White Dwarf Candidates Formed Through Binary evolution in Star Clusters}

\correspondingauthor{Chengyuan Li}
\email{lichengy5@mail.sysu.edu.cn}

\author[0009-0005-8565-0858]{Huahui Yan}
\affiliation{School of Physics and Astronomy, Sun Yat-sen University, Daxue Road, Zhuhai 519082, People's Republic of China.}
\affiliation{CSST Science Center for the Guangdong--Hong Kong--Macau Greater Bay Area, Zhuhai 519082, People's Republic of China}

\author[0000-0003-3471-9489]{Li Wang}
\affiliation{School of Physics and Astronomy, Sun Yat-sen University, Daxue Road, Zhuhai 519082, People's Republic of China.}
\affiliation{CSST Science Center for the Guangdong--Hong Kong--Macau Greater Bay Area, Zhuhai 519082, People's Republic of China}

\author[0000-0002-4591-1903]{David R. Miller}
\affiliation{Department of Physics and Astronomy, University of British Columbia, Vancouver, BC V6T 1Z1, Canada}

\author{Chenyu He}
\affiliation{School of Physics and Astronomy, Sun Yat-sen University, Daxue Road, Zhuhai 519082, People's Republic of China.}
\affiliation{CSST Science Center for the Guangdong--Hong Kong--Macau Greater Bay Area, Zhuhai 519082, People's Republic of China}

\author{Jiamao Lin}
\affiliation{School of Physics and Astronomy, Sun Yat-sen University, Daxue Road, Zhuhai 519082, People's Republic of China.}
\affiliation{CSST Science Center for the Guangdong--Hong Kong--Macau Greater Bay Area, Zhuhai 519082, People's Republic of China}

\author[0000-0003-3389-2263]{Xiaoying Pang}
\affiliation{Department of Physics, Xi'an Jiaotong--Liverpool University, 111 Ren’ai Road, Dushu Lake Science and Education Innovation District, Suzhou 215123, Jiangsu Province, P.R. China.}
\affiliation{Shanghai Key Laboratory for Astrophysics, Shanghai Normal University, 
                100 Guilin Road, Shanghai 200234, P.R. China}

\author{Jingkun Zhao}
\affiliation{National Astronomical Observatories, Chinese Academy of Sciences, Beijing 100101, PR China}

\author{Jincheng Guo}
\affiliation{Department of Scientific Research, Beijing Planetarium, Xizhimenwai Street, Beijing 100044, PR China}

\author[0000-0002-7203-5996]{Richard de Grijs}

\affiliation{School of Mathematical and Physical Sciences, Macquarie University, Balaclava Road, Sydney NSW 2109, Australia}

\affiliation{Astrophysics and Space Technologies Research Centre, Macquarie University, Balaclava Road, Sydney, NSW 2109, Australia}

\affiliation{International Space Science Institute--Beijing, 1 Nanertiao, Zhongguancun, Hai Dian District, Beijing 100190, China}

\author[0000-0002-6398-0195]{Hongwei Ge}

\affiliation{International Centre of Supernovae (ICESUN), Yunnan  Key Laboratory of 
Supernova Research, Yunnan Observatories, Chinese Academy of Sciences (CAS), 
Kunming 650216, People’s Republic of China} 
\affiliation{University of Chinese Academy of Sciences, Beijing 100049, People's 
Republic of China}

\author{Zhen Guo}
\affiliation{Instituto de F{\'i}sica y Astronom{\'i}a, Universidad de Valpara{\'i}so, ave. Gran Breta{\~n}a, 1111, Casilla 5030, Valpara{\'i}so, Chile}

\author{Bo Ma}
\affiliation{School of Physics and Astronomy, Sun Yat-sen University, Daxue Road, Zhuhai 519082, People's Republic of China.}
\affiliation{CSST Science Center for the Guangdong--Hong Kong--Macau Greater Bay Area, Zhuhai 519082, People's Republic of China}

\author{Dichang Chen}
\affiliation{School of Physics and Astronomy, Sun Yat-sen University, Daxue Road, Zhuhai 519082, People's Republic of China.}
\affiliation{CSST Science Center for the Guangdong--Hong Kong--Macau Greater Bay Area, Zhuhai 519082, People's Republic of China}



\author{Chengyuan Li}
\affiliation{School of Physics and Astronomy, Sun Yat-sen University, Daxue Road, Zhuhai 519082, People's Republic of China.}
\affiliation{CSST Science Center for the Guangdong–Hong Kong–Macau Greater Bay Area, Zhuhai 519082, People's Republic of China}








\begin{abstract}
White dwarfs (WDs), the evolutionary endpoints of most stars, can form through both single-star and binary channels. While single-star evolutionary models enable reliable WD age estimates, binary evolution introduces interactions that can accelerate WD formation and result in a variety of exotic WDs, which may exhibit strong magnetic fields, rapid rotation, or even serve as potential gravitational wave sources. Such systems offer valuable insights into magnetic field generation, angular momentum evolution, and compact object physics. Star clusters, with their approximately coeval populations, allow precise age determination of member WDs. If a WD's total age derived from single-star evolution exceeds that of its host cluster, it likely indicates a binary origin. In this study, we use \textit{Gaia} 5D astrometry to identify 439 WD candidates in 117 open clusters, with 244 likely formed via binary evolution. We discuss the possibility of dynamical ejection for WDs meeting only 2D (proper motion space) membership criteria.
Spectroscopic observations further reveal a subset with strong magnetic fields and rapid rotation, supporting their binary evolutionary origin.
\end{abstract}


\section{Introduction} \label{sec:intro}
White dwarfs (WDs) are compact stellar remnants formed from stars with initial masses below about 8–10 $M_{\odot}$ \citep{Doherty2015MNRAS.446.2599D}. With radii comparable to the Earth's and masses similar to the Sun's, these objects occupy the faint and blue region of the color--magnitude diagram (CMD). As WDs no longer sustain nuclear fusion, they gradually cool over astronomical timescales. Approximately 97\% of stars in the Galaxy will ultimately evolve into WDs \citep{Woosley2015, Lauffer2018}, making them essential for understanding stellar evolution. 

WDs are uniquely suited for detailed spectroscopic and photometric studies because their observable Ultraviolet/optical signals directly reflect intrinsic thermal emission. Based on their spectral features, approximately 80\% of WDs exhibit hydrogen-dominated absorption lines and are classified as DA WDs. DA WDs may exhibit pulsations (known as DAVs or ZZ Ceti stars) within the effective temperature range of 10,500–13,000 K \citep{Hermes2017ApJS..232...23H, Lasker1971ApJ...163L..89L}. Besides the DA type, other main spectral classes of WDs include DB, DC, DQ, and DZ. DB WDs are characterized by helium absorption lines, and DB WDs pulsate ( referred to as DBVs/V777 Her stars) at temperatures between 22,000 and 31,000 K \citep{Vanderbosch2022ApJ...927..158V}. DC WDs exhibit featureless spectra owing to their sufficiently low effective temperatures that prevent atomic excitation and subsequent absorption line features. During the late stages of WD cooling evolution, decreasing atmospheric temperatures cause the gradual disappearance of hydrogen (DA) and helium (DB) spectral features, resulting in a transition to the featureless DC type. The presence of strong magnetic fields in some DA and DB WDs can lead to spectra that resemble the featureless DC type. This occurs due to extreme Zeeman splitting, which distorts and broadens absorption lines to such an extent that they become completely blended. Consequently, although hydrogen or helium is present in the atmosphere, all characteristic spectral features are indistinguishable, resulting in an apparently smooth and continuous spectrum \citep{Kilic2025ApJ...979..157K}. DQ WDs display molecular carbon absorption bands in their spectra. Despite the similar nomenclature, it is important to note that the hot DQ WDs represent a distinct physical class rather than merely a subset of classical DQs. These hot, carbon-rich WDs are hypothesized to originate from double WD (DWD) mergers, as their atmospheric composition and thermal properties are inconsistent with single-star evolution\citep{Kawka2023MNRAS.520.6299K}.
DZ WDs exhibit prominent metal absorption lines in their spectra, with atmospheric pollution generally attributed to external accretion rather than primordial composition. The metal enrichment (commonly termed pollution) arises from the accretion of both planetary system remnants (e.g., asteroids, comets) and stellar evolutionary byproducts (e.g., circumstellar material), which are dynamically perturbed into star-grazing orbits before being accreted onto the WD surface. The observed metal lines in WDs require ongoing accretion, as their short diffusion timescales ($\sim10^6$ yr) compared to cooling ages ($\sim10^9$ yr) imply that single accretion events would become undetectable before observation \citep{Pelisoli2025arXiv250219496P}.

WD binaries are important for the study of stellar evolution and multi-messenger astronomy. They are widely recognized as predecessors of Type Ia supernovae (SNe Ia). In the single-degenerate scenario, the WD continuously accretes material from a non-degenerate companion, causing it to exceed its Chandrasekhar limit thereby triggering a thermonuclear explosion \citep{Nomoto1984ApJ...286..644N}. In the double-degenerate scenario, the merger of two carbon-oxygen core WDs results in a combined mass that exceeds the Chandrasekhar limit, thereby producing a SNe Ia explosion \citep{Webbink1984ApJ...277..355W}. Numerical simulations show that, given specific accretion rates, a SNe Ia explosion may also be triggered at a sub-Chandrasekhar mass \citep{Woosley2011ApJ...734...38W}. Another example is short-period close WD binaries (CWDBs). CWDBs with orbital periods shorter than 60 minutes emit powerful gravitational waves in the millihertz band, expected to be detectable by space-based gravitational wave observatories such as the Laser Interferometer Space Antenna (LISA) \citep{Amaro-Seoane2017arXiv170200786A} and TianQin \citep{Luo2016CQGra..33c5010L}. Population synthesis models predict the Milky Way should contain hundreds of millions of double white dwarf (DWD) systems \citep{Nelemans2001A&A...365..491N, Korol2022MNRAS.511.5936K}. However, current observations have identified only about 300 DWD systems, highlighting a significant gap between theoretical predictions and observational capabilities owing to their intrinsic faintness.

Identification of WD binaries can be achieved through both spectroscopic analysis and time-domain photometric variability studies. \cite{Yan2023Univ....9..177Y,Yan2024A&A...684A.103Y} used spectral data to search for WD binary candidates by measuring the radial velocity (RV) variations. \cite{Ren2023ApJS..264...39R} searched for about 400 short-period CWDB candidates based on time-domain photometry. Within the local sample, approximately 50\% of intermediate-mass (1.5--5 M$_\odot$) main-sequence (MS) stars exist in binary systems, a binary fraction twice as high as that observed among WDs \citep{Ferrario2012MNRAS.426.2500F}. \cite{Ferrario2012MNRAS.426.2500F} proposed that this discrepancy arises because many WDs are obscured by their brighter companions and thus excluded from detection. \cite{Toonen2017A&A...602A..16T} proposed that the reduced binary fraction among WDs primarily results from post-MS merger events. Mergers can occur throughout WD binary evolution, potentially producing rapidly rotating, highly magnetic, and massive WDs. 

The detection of extremely low-mass WDs (ELM WDs; $<0.3 M_{\odot}$, helium-core) provides an efficient method for identifying WD binaries. Due to their insufficient mass to undergo single-star evolution within a Hubble time, ELM WDs must originate from binary interactions. Indeed, observations confirm that most ELM WDs reside in binary systems. \citep{Brown2013ApJ...769...66B,Brown2016ApJ...818..155B,Kilic2011ApJ...727....3K,Kilic2012ApJ...751..141K}. Similarly, precise age determinations of WDs enable tests of their formation channels by comparing their measured ages with the timescales required for single-star evolution. A WD's total age (cooling age + progenitor lifetime) determination falling below the theoretically required formation time for single-star evolution may suggest a binary origin. Star clusters, particularly open clusters (OCs), represent single stellar populations where all members formed from the same molecular cloud \citep{BruzualA2010RSPTA.368..783B} and consequently share nearly identical ages. Therefore, cluster membership enables precise age determination of WDs, providing critical constraints for investigating their formation and evolutionary timescales.

Before the  \textit{Gaia} era, studies of WDs relied primarily on spectroscopic surveys. Thanks to large-scale spectroscopic surveys such as the Large Sky Area Multi-Object Fiber Spectroscopic Telescope (LAMOST) \citep{Zhao2012RAA....12..723Z, Zhao2013AJ....145..169Z, Guo2022MNRAS.509.2674G} and the Sloan Digital Sky Survey (SDSS) \citep{York2000AJ....120.1579Y, Kepler2019}, tens of thousands of WDs have been discovered.  \textit{Gaia}'s high-precision astrometry has revolutionized WD searches by providing unprecedented parallax and proper motion measurements, along with photometric data in the $G$, $G_{\text{BP}}$, and $G_{\text{RP}}$ bands, significantly improving detection efficiency \citep{GaiaCollaboration2018A&A...616A...1G, GaiaCollaboration2021A&A...649A...1G}. In particular, Gaia Data Release 3 (DR3) has led to the discovery of about 1.3 million WD candidates, including 359,000 high-confidence WD candidates \citep{2021MNRASGentileFusillo}. This large WD sample provides an unprecedented opportunity to understand the formation and evolution of WDs \citep{Tremblay2024NewAR..9901705T}. 

The availability of  \textit{Gaia}'s precise astrometric data has also led to a significant increase in OC discoveries. Cluster identification is achieved by detecting stellar aggregates sharing common astrometric parameters (position, parallax, and proper motion). Several studies have significantly expanded the OC census through  \textit{Gaia} data using clustering algorithms, including \cite{CastroGinard2018A&A...618A..59C,CastroGinard2019A&A...627A..35C,CastroGinard2020A&A...635A..45C}, \cite{Liupang2019ApJS..245...32L}, \cite{Cantat-Gaudin2018A&A...618A..93C}, and \cite{Hunt2021A&A...646A.104H,Hunt2023A&A...673A.114H,Hunt2024A&A...686A..42H}. The growing samples of WDs and OCs have significantly advanced studies of WD populations in clusters. For example, while the theoretical WD mass limit is well established at 1.38 $M_{\odot}$ \citep{Nomoto1987ApJ...322..206N}, the progenitor mass limit remains debated \citep{Weidemann1983A&A...121...77W, Horiuchi2011ApJ...738..154H, Kroupa2003ApJ...598.1076K}.
The upper mass limit can be constrained by examining the initial-final mass relation (IFMR). While this relation can be explored through various means, OCs provide a particularly direct environment for such studies, allowing for precise constraints on the high-mass end of the relation. This cluster-based approach has been extensively utilized in numerous investigations, including \cite{Kalirai2008ApJ...676..594K, Williams2009ApJ...693..355W, Cummings2018ApJ...866...21C, Prisegen2021A&A...645A..13P, Prisegen2023A&A...678A..20P}. A notable discrepancy is the observed deficit of WDs in OCs compared to theoretical predictions. This suggests that natal kicks during WD formation impart sufficient velocities to gradually eject them from cluster potentials \citep{Fellhauer2003ApJ...595L..53F,Kalirai2008ApJ...676..594K}. \cite{Miller2022ApJ...926L..24M,Miller2023ApJ...956L..41M} conducted a search for escaped massive WDs in clusters based on kinematic retrieval and identified one of the most massive WDs known to have evolved through a single-star channel, with a mass of 1.317 $M_{\odot}$. Common envelope (CE) evolution remains one of the most uncertain phases in binary stellar evolution \citep{Webbink2008ASSL..352..233W,Chen2024PrPNP.13404083C,Ge2024arXiv241117333G}, primarily due to computational complexities and the scarcity of observed systems with both pre- and post-CE constraints. Studying post-CE systems in OCs offers a partial solution by providing independent age determinations. \cite{Grondin2024arXiv240704775G} systematically identified 52 high-probability WD--MS binaries across 38 OCs. This sample enables crucial connections between post-CE systems and their progenitor configurations.

In this study, we systematically identify WDs in OCs that may have undergone binary interactions,  employing  \textit{Gaia} astrometric parameters (positions, parallaxes, and proper motions) to establish robust cluster--WD associations. WDs with theoretical single-star formation timescales (cooling age + progenitor age) significantly longer than the age of their host clusters are identified as binary evolution candidates. We compiled archival spectroscopy and photometry for some selected candidates, revealing strong magnetic fields and rapid rotation—possible signatures of binary evolution.

This paper is organized as follows. Section \ref{sec2} describes our data sources and method for selecting WD members in OCs. Section \ref{sec3} presents the derivation of WD parameters, including age, mass, effective temperature, and surface gravity, through model calculations. Section \ref{sec4} reports our main findings, while Section \ref{sec5} briefly discusses the escape mechanism driven by natal kicks for WDs with consistent proper motions. Finally, we summarize our conclusions in Section \ref{sec6}.

\section{Data selection}\label{sec2}
\subsection{White dwarf sample}
\cite{Gentile2019MNRAS.482.4570G} screened 486,641 WD candidates using the  \textit{Gaia} DR2 catalog, with about 260,000 high-confidence WD candidates in the magnitude range $8 < G < 21$ mag. They first performed an initial cut of the  \textit{Gaia} data based on the range of WDs confirmed in the SDSS spectra in the CMD, then removed objects with unreliable measurements based on  \textit{Gaia} quality flags\footnote{Specifically, \texttt{PHOT\_BP\_RP\_EXCESS\_FACTOR} was used to filter objects with unreliable colors or a bright sky background, \texttt{ASTROMETRIC\_EXCESS\_NOISE} was used to identify objects with reliable parallax measurements, \texttt{ASTROMETRIC\_SIGMA5D\_MAX} was used to filter out objects where the  \textit{Gaia} parameter had particularly poor precision in either dimension}. Subsequently, they excised targets belonging to the Galactic disk and the Magellanic Clouds to produce a relatively clean and reliable sample of WD candidates from  \textit{Gaia} DR2 (hereafter, the GDR2 WD sample). Using  \textit{Gaia} Early Date Release 3 (EDR3), \cite{2021MNRASGentileFusillo} identified approximately 1.3 million WD candidates, including roughly 359,000 high-confidence objects (hereafter referred to as the GDR3 WD sample). Their selection methodology closely follows the WD screening approach presented in \cite{Gentile2019MNRAS.482.4570G}.

By directly comparing the targets with unique  \textit{Gaia} source IDs in the GDR2 and GDR3 WD samples, 9,726 DR2 WD candidates were completely excluded from the DR3 WD sample of  \cite{2021MNRASGentileFusillo}. These targets were eliminated from the sample as their photometric and astrometric measurements in GDR3 were deemed unreliable as WD selection criteria. In our study, we used all WD candidates from both the GDR2 and GDR3 samples for OC matching. However, we primarily present the results derived from the GDR3 sample, while those based on GDR2 are provided in Appendix \ref{Appendix A}, Table \ref{tab:dr2wd}, and Figure Set 1. For the GDR2 WD sample, we performed a cross-match using the updated parameter measurements from Gaia DR3.

\subsection{Open cluster sample} 
Using a blind all-sky search of 729 million sources from  \textit{Gaia} DR3 with magnitudes brighter than $G\approx 20$ magnitudes, \cite{Hunt2023A&A...673A.114H} constructed a large homogenized catalog of OCs, which contains a large number of newly discovered clusters. They recovered 7,167 clusters, of which 2,387 are newly discovered. However, the potential lack of gravitational binding in many of these clusters makes the OC sample less reliable for certain scientific applications. To address this issue, \cite{Hunt2024A&A...686A..42H} performed a reanalysis of their previous catalog \cite{Hunt2023A&A...673A.114H}. By excluding globular clusters and sources located beyond 15 kpc, they derived complementary photometric mass estimates (corrected for systematic biases) for 6,956 clusters. Using these masses, they computed each cluster’s Jacobi radius (Roche surface size) to differentiate gravitationally bound from unbound systems. This yielded a final sample of 5,647 bound clusters. From these, a high-quality subset of 3,530 clusters was selected based on two criteria: 

\begin{itemize}
\item[1.]Clustering Significance Test (CST): Evaluates the density contrast between cluster candidates and field stars using a nearest neighbor approach.
\item[2.]CMD Class: Assessing the likelihood of a cluster being a single, coeval stellar population.
\end{itemize}

We restricted our sample to OCs located within 1.1 kpc of the Sun. The distance measurements are based on the OCs parallax provided in the \cite{Hunt2024A&A...686A..42H} catalog. Beyond this distance, the increasing astrometric uncertainties in  \textit{Gaia} data, compounded by the intrinsic faintness of WDs, would significantly reduce the reliability of OC–WD pair identifications. From an initial parent sample of 3,530 high-quality OCs, we identified a subset of 752 systems satisfying this distance criterion.

\subsection{Identifying white dwarfs in open clusters}

We began by applying broad matching criteria in positional and kinematic parameter space to identify initial candidate WD--OC associations. Following \cite{Prisegen2021A&A...645A..13P} and \cite{Prisegen2023A&A...678A..20P}, we apply a spatial selection criterion of $\theta < 5 \times r_{50}$, where $\theta$ is the WD-OC center angular separation and $r_{50}$ denotes the half-member radius from \cite{Hunt2024A&A...686A..42H}. Next, for the parallax and kinematic matching, similarly to \cite{Grondin2024arXiv240704775G}, the filtering conditions are as follows:
\begin{equation}
\left( \pi, \mu_{\alpha}, \mu_{\delta} \right) < 1.6\times P .
\end{equation}
The parameter $\pi$ denotes parallax, while $\mu_{\alpha}$ and $\mu_{\delta}$ represent the proper motion components in right ascension and declination, respectively. The symbol $P$ refers to the range of astrometric parameter values spanned by members of a given OC. For example, for a given OC, we have the parallax and proper motion information of its member stars. When selecting WDs within the OC, we choose a larger range, which is 1.6 times the difference between the maximum and minimum values of the parallax and proper motion parameters of the OC's member stars. After the initial rough screening of the positional and kinematic parameters, we obtained a preliminary sample of WDs corresponding to their parent OCs.

Subsequently, we applied more stringent selection criteria to the initially matched sample to refine the candidate WD–OC associations. We adopted the approach established by \cite{O'Grady2020ApJ...901..135O}, \cite{Neugent2020ApJ...900..118N}, and \cite{O'Grady2023ApJ...943...18O}. This method has recently been successfully applied by \cite{Grondin2024arXiv240704775G} in the search for WD--MS binary systems within OCs. The main steps of the procedure are as follows: For each OC, two independent covariance matrices (denoted as 
$C$) were constructed based on the astrometric parameters from  \textit{Gaia} DR3.  The first matrix is the 2\,D kinematic distribution of OC members, where the components of the vector 
$\boldsymbol{\mu}$ represent the proper motions in right ascension ($\mu_\alpha$) and declination ($\mu_\delta$):
\begin{equation}
\boldsymbol{\mu} = [\mu_\alpha, \mu_\delta ]
\end{equation}
The covariance matrix $C$ is constructed to describe the relationships between these components:
\begin{equation}
C = \begin{vmatrix}
c_{\mu_\alpha\mu_\alpha} & c_{\mu_\alpha\mu_\delta} \\ 
c_{\mu_\delta\mu_\alpha} & c_{\mu_\delta\mu_\delta}  \\ \end{vmatrix}.
\end{equation}
Here, $c_{\mu_\alpha\mu_\alpha}$, $c_{\mu_\alpha\mu_\delta}$ and other elements represent the covariance between the proper motion components, providing insights into how they co-vary. The second matrix represents the 3\,D spatial distribution of OC members, with the vector
$\boldsymbol{\mu}$ now consisting of the right ascension ($\alpha$), declination ($\delta$), and parallax ($\pi$) of the stars. The covariance matrix $C$ is simillar to the 2\,D covariance matrix:
\begin{equation}
\boldsymbol{\mu} = [\alpha, \delta, \pi]
\end{equation}
\begin{equation}
C = \begin{vmatrix}
c_{\alpha\alpha} & c_{\alpha\delta} & c_{\alpha\pi} \\ c_{\delta\alpha} & c_{\delta\delta} & c_{\delta\pi}  \\ c_{\pi\alpha} & c_{\pi\delta} & c_{\pi\pi} \end{vmatrix}.
\end{equation}

We use the 2D kinematic distribution as an example to demonstrate the membership selection process.  For the cluster members, after accounting for the individual measurement uncertainties $\sigma_{\mu_\alpha,i}$, $\sigma_{\mu_\delta,i}$, from each OC member 
$i$, the total likelihood across all OC members is expressed as: 
\begin{align}
\ln L &= -\frac{1}{2} \sum_{i} \left( \boldsymbol{\mu}_i - \boldsymbol{\mu} \right)^T (C_{\text{tot},i})^{-1} \left( \boldsymbol{\mu}_i - \boldsymbol{\mu} \right) \notag \\
&\quad + \ln \left( \det(2\pi C_{\text{tot},i}) \right).
\end{align}
where $\boldsymbol{\mu_{i}} = [\mu_{\alpha i}, \mu_{\delta i} ]$ represents the proper motion of OC member $i$, $C_{\text{tot},i}$ is the effective total covariance for OC member $i$ and  
\begin{equation}
C_{\text{tot},i} = C +  
\begin{bmatrix}  
\sigma_{\mu_{\alpha,i}}^2 & 0 \\  
0 & \sigma_{\mu_{\delta, i}}^2 
\end{bmatrix}.
\end{equation}

The method intrinsically weighs object contributions according to their measurement uncertainties, with larger uncertainties generating correspondingly larger covariance values. We employ \texttt{scipy.optimize.minimize}\footnote{\url{https://scipy.org/}} to minimize the total negative log-likelihood, allowing us to find the optimal parameter solution. We denote the estimated optimal covariance as $\boldsymbol{C_{*}}$, to distinguish it from the covariance $\boldsymbol{C}$. For 3\,D spatial distributions, the computational procedure follows the same methodology as described for the 2\,D kinematic case.

Finally, for each WD obtained from the initial screening, we calculate its $\chi^2$ value in the 2\,D kinematic and 3\,D spatial distributions, respectively. The $\chi^2$ is calculated as: 
\begin{equation}
    \chi^2 = (\boldsymbol{\mu} - \boldsymbol{X})^T \boldsymbol{C}_{*}^{-1}(\boldsymbol{\mu} - \boldsymbol{X}).
\end{equation}
where $\boldsymbol{X}$ represents the median spatial or kinematic properties of OC members. To determine if a WD is spatially or kinematically consistent with a particular cluster, we establish thresholds of $\chi^2<12.8$ for our 3\,D spatial analysis and $\chi^2<10.6$ for our 2\,D kinematic analysis. For the  $\chi^2$ statistic, there is a 99.5\% probability that these two thresholds are satisfied and are plausible \citep{Grondin2024arXiv240704775G}.

\begin{figure}[ht!]
\centering
\includegraphics[width=0.5\textwidth]{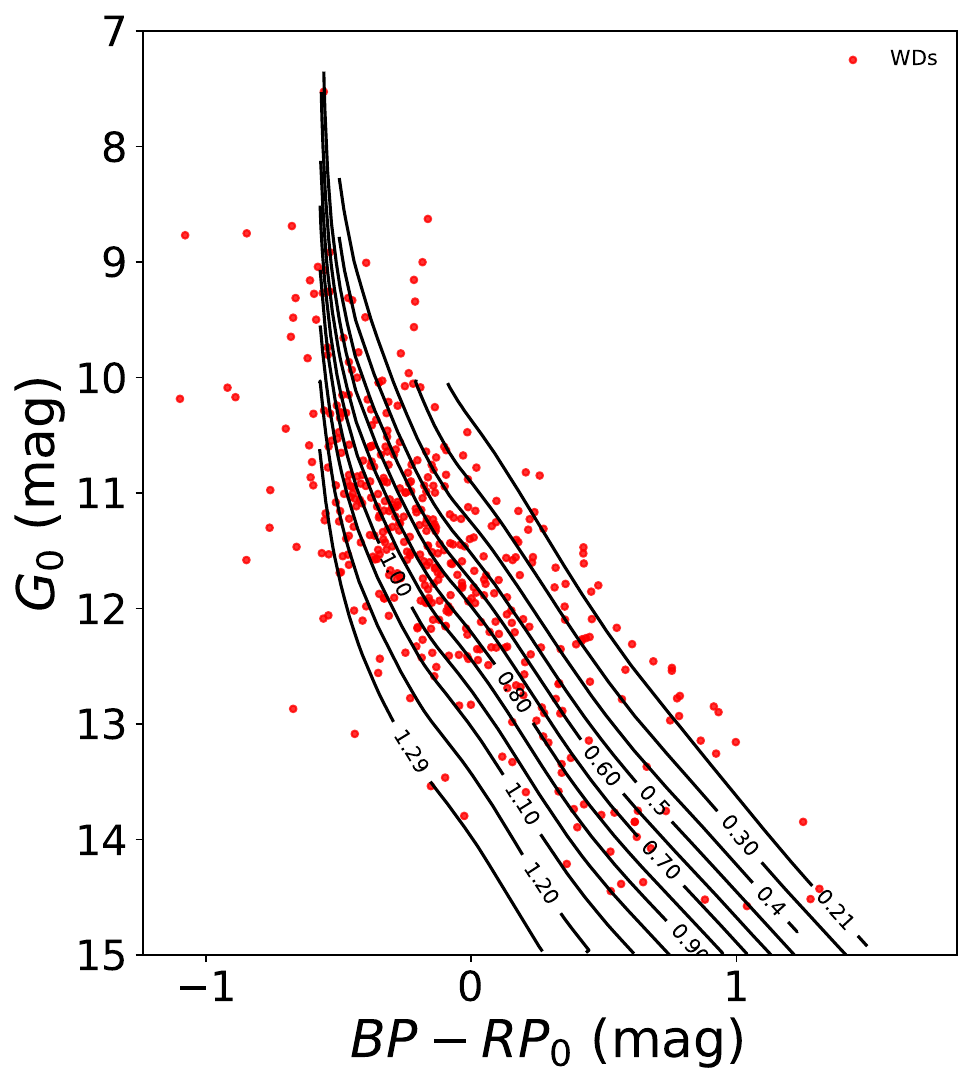}
\caption{The distribution of WDs belonging to OCs on the CMD. Primarily vertical tracks show mass models \citep{Bedard2020ApJ...901...93B} from 0.21 $M_{\odot}$ to 1.29 $M_{\odot}$ . 
\label{fig:1}}
\end{figure}

After applying our selection criteria, we compare the spatial (equatorial coordinates), kinematic (proper motion), and parallax distributions of our screened WDs with cluster members from \cite{Hunt2024A&A...686A..42H}. Targets that did not agree with the cluster members' distribution in any sector of parameter space were removed by visual inspection. From our analysis, we identified 439 WD candidates across 117 clusters. We classified these candidates using the following labeling scheme: `2D' for objects satisfying only the 2D membership screening condition (proper motion), `3D' for those meeting only the 3D membership criteria (equatorial coordinates and parallax), and `5D' for candidates fulfilling both 2D and 3D membership screening conditions. It should be noted that while the 2D and 3D candidates have lower reliability than the 5D sample because they meet membership requirements in fewer dimensions, they remain plausible members. These candidates may represent escaping stars driven by natal kicks or binary evolution that no longer share the cluster's tight kinematic core. Figure \ref{fig:1} presents the distribution of these selected WDs in the CMD.
 
\section{Parameter estimation}\label{sec3}
We determined the ages of the OCs in our sample by fitting isochrones from the PARSEC version 1.2S models\citep{Bressan2012MNRAS.427..127B, Tang2014MNRAS.445.4287T, Nguyen2022A&A...665A.126N, Bohlin2020AJ....160...21B}\footnote{\url{https://stev.oapd.inaf.it/cgi-bin/cmd}}. In our fitting procedure, we assumed solar metallicity ($Z = 0.015$) and adopted an age resolution of $\Delta \log (t/\mathrm{yr}) = 0.05$ for the isochrone grid.

We employed the \texttt{WD\_models} Python package\footnote{A Python package for transforming WD photometry and deriving physical parameters based on WD models. \url{https://github.com/SihaoCheng/WD\_models}} to estimate physical parameters from the extinction-corrected  \textit{Gaia} photometry. 
This package provides a grid of model atmospheres for WDs; since the majority of WDs are DA type, we used pure-hydrogen atmosphere models with carbon–oxygen cores \citep{Bedard2020ApJ...901...93B}\footnote{\url{http://www.astro.umontreal.ca/~bergeron/CoolingModels}}. We adopted the dereddened photometry from \cite{2021MNRASGentileFusillo}, who derived individual extinction estimates using a local 3D dust map. Using these extinction-corrected $\sl Gaia $ magnitudes ($A_G = 0.835A_V$, $A_{BP} = 1.139A_V$, $A_{RP} = 0.650A_V$), we derived fundamental WD parameters, including mass ($M_\odot$), effective temperature ($T_{\mathrm{eff}}$), surface gravity ($\log {g}$), cooling age, and total age\footnote{Although the \citet{2021MNRASGentileFusillo} catalog provides some WD parameters, we recalculated them to ensure the consistency of our data processing pipeline.}. The total age is the sum of the cooling age and the progenitor's MS lifetime.
To compute the total age of a WD, \texttt{WD\_models} employ an iterative procedure that combines the initial–final mass relation from \cite{Cummings2018ApJ...866...21C} with MESA Isochrones and Stellar Tracks (MIST) stellar evolution tracks \citep{Choi2016ApJ...823..102C}. This enables the estimation of the total age for WDs under the assumption of single-star evolution. However, we explicitly caution that these derived total ages are subject to significant uncertainties and should be regarded as indicative estimates rather than precise determinations. 
To quantify the uncertainties of the derived parameters, we employed a Monte Carlo method. For each WD, we generated 1,000 synthetic samples by resampling the photometry from Gaussian distributions centered on the observed magnitudes, with widths corresponding to their measurement errors. We then derived the physical parameters for each of these 1,000 simulated sets. Finally, the standard deviation of the resulting parameter distribution was adopted as the uncertainty estimate. The photometric uncertainty in $(BP - RP)$ parameter was estimated through error propagation using the following equation:
\begin{equation}
    \sigma_{\text{BP-RP}} = \sqrt{\sigma_{\text{BP}}^2 + \sigma_{\text{RP}}^2}
\end{equation}
$\sigma_{\text{BP}}, \sigma_{\text{RP}}$ represent the observational errors in the respective bands. We used:
\begin{equation}
    G_{\text{abs}} = G + 5 \times \left( \log_{10}(\pi\times 0.001) + 1 \right)
\end{equation}
to convert the absolute magnitude. For the absolute $G$ magnitude error, the formula for the converted absolute $G$ magnitude is:
\begin{equation}
    \sigma_{G_{\text{abs}}} = \sqrt{\sigma_{G}^{2} + \left( \frac{5}{\ln(10)} \cdot \frac{\sigma_{Plx}}{Plx} \right)^{2}}.
\end{equation}

To systematically identify WDs that likely originated from binary evolutionary channels, we implemented a screening procedure based on age consistency. We employed a Monte Carlo approach to estimate the probability that a WD could not have formed via single-star evolution given the age constraints of its host OC. For each WD, we generated 1,000 samples drawn from a Gaussian distribution centered on the measured mass, with the standard deviation set to the associated mass error. We then determined the fraction of these samples whose derived total ages exceeded the OC's age, interpreting this fraction as the probability of a binary origin ($P_{\text{bin}}$). WDs exhibiting $P_{\text{bin}}>0.8$ are considered likely to be either currently in binary systems or to have experienced binary evolution. We note that targets near the boundary of the WD cooling model, those with low WD probabilities ($P_{\rm wd}$) from \cite{2021MNRASGentileFusillo}, or those with large photometric errors may yield less reliable probability estimates. However, for systems where $P_{\text{bin}} > 0.8$, the significant discrepancy between their formation timescales and the cluster age persists even within error bounds, providing robust evidence for binary evolution.

\section{Main Results}\label{sec4}
We have identified 439 WDs potentially associated with 117 OCs. Our classification reveals:
\begin{itemize}
\item 219 WDs satisfying only the 2D membership screening criterion
\item 109 WDs satisfying only the 3D membership screening criterion
\item 111 WDs satisfying both criteria (5D membership screening criterion)
\end{itemize}

We find 243 WDs with high probability ($P_{\mathrm{bin}} > 0.8$) of having undergone binary evolution, suggesting their formation may have been influenced by binary interactions. Note that in calculating the probability of formation through binary evolution, we performed 1,000 Monte Carlo simulations for each target to account for photometric errors. Any simulated points falling outside the valid range of the WD cooling models, where physical parameters could not be determined, were classified as non-binary evolution instances for conservative estimation. Consequently, probabilities for targets near model boundaries may be underestimated, leading to their exclusion.  This subset comprises:
\begin{itemize}
\item 123 WDs satisfying only 2D criteria
\item 72 WDs satisfying only 3D criteria
\item 48 WDs satisfying both criteria
\end{itemize}

These binary-evolved WD candidates show distinct age distributions:
\begin{itemize}
\item 33 WDs in extremely young OCs ($\leq$25 Myr)
\item 105 WDs in young OCs (25--100 Myr)
\item 102 WDs in intermediate-age OCs (100--1000 Myr)
\item 3 WDs in old OCs ($>$1000 Myr)
\end{itemize}

\begin{figure*}[ht!]
\centering
\includegraphics[width=1.0\textwidth]{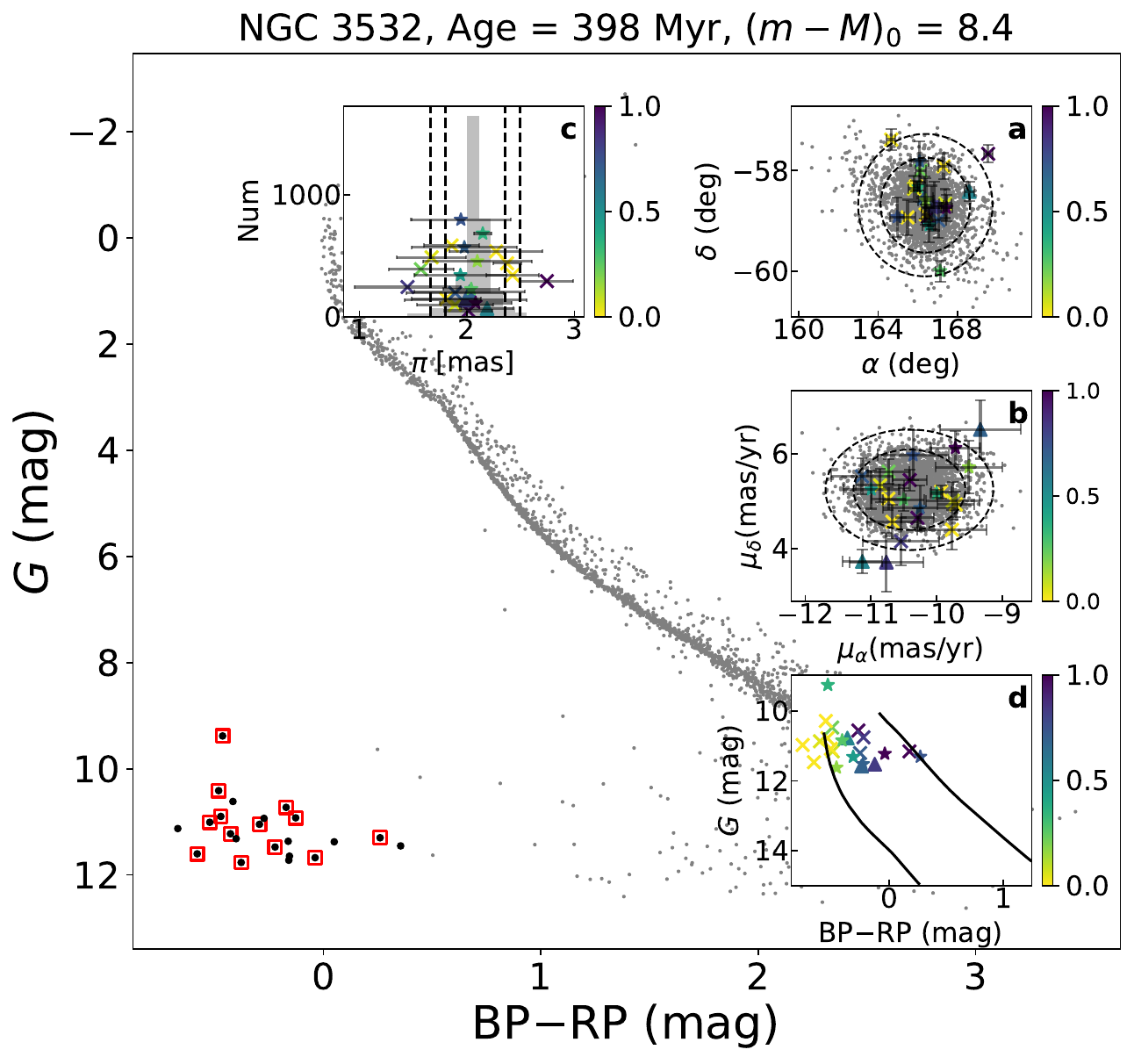}
\caption{The spatial distribution, kinematic parameters, and CMD of candidate cluster members from \cite{Hunt2024A&A...686A..42H}, along with the selected WD candidates. The main plot is the CMD of the cluster, in which gray points represent the OC member stars selected by \cite{Hunt2024A&A...686A..42H}, while black points indicate the WDs we identified as belonging to the cluster. Red squares denote WDs previously identified in the literature as cluster members. Inset (a): Spatial distribution of cluster members and WDs.  
Inset (b): Proper motion distribution of cluster members and WDs.  
Inset (c): Parallax distribution of cluster members and WDs.  Inset (d): CMD for WD members in cluster. The tracks represent mass models with 0.21 $M_{\odot}$ and 1.29 $M_{\odot}$. In each inset, black dashed lines indicate the $1\sigma$ and $3\sigma$ ranges. Crosses, triangles, and pentagrams denote WDs satisfying the 2D, 3D, and 5D selection criteria, respectively. The color bar represents the probability of a binary origin.
\label{fig2}}
\end{figure*}

As an example, Figure \ref{fig2} illustrates the spatial and kinematic distributions, along with the CMD, of candidate members in NGC 3532, including its WD members. The color bar indicates the probability of a binary origin, with red squares denoting previously identified candidate WD members. Similar results for other OCs are presented in Appendix \ref{appendix B}. Table \ref{tab:oc_para} shows properties of OCs with at least one associated WD candidate. Tables \ref{tab:elmocwd}, \ref{tab:yocwd}, \ref{tab:mocwd}, and \ref{tab:oocwd} list the WD members recovered in extremely young, young, intermediate-age, and older OCs, respectively.

For each cluster, we use the initial mass function \cite[IMF,][]{Kroupa2001MNRAS.322..231K, Kroupa2002Sci...295...82K}, combined with the PARSEC model to calculate the expected number of WDs. We used the same method as applied in \cite{Richer2021ApJ...912..165R}. The equation for the expected number of WDs is given in:
\begin{equation}
\langle N_{\text{WD}} \rangle = N_1 \left[ \int_{M_{\text{init,3}}}^{M_{\text{TO}}} \frac{\text{d}N}{\text{d}M} \text{d}M \right]^{-1} \int_{M_{\text{init, WD}}}^{12M_{\odot}} \frac{\text{d}N}{\text{d}M} \text{d}M,
\end{equation}
where $N_1$ is the number of stars in the top one-third of the brightest MS stars in the cluster. $M_{\text{init,3}}$ is the initial mass of the faintest of the $N_1$ stars. $M_{\text{TO}}$ is the turnoff mass of the cluster. $M_{\text{init, WD}}$ is the initial mass of a star that would produce a WD at the present time, and the upper mass limit for WD production is set to $12M_{\odot}$. This number is used to compare with the detected number of WDs in clusters, thus giving an estimate of the extent to which WDs in the star cluster have been completely detected.


\subsection{Extremely young open clusters}
We classified OCs younger than 25 Myr as extremely young OCs, where WD formation via single-star evolution is highly unlikely \citep{Limongi2024ApJS..270...29L}. The MS turnoff mass at this age is approximately 10$M_{\odot}$, corresponding to the theoretical upper mass limit for WD production \citep{Kroupa2003ApJ...598.1076K}. In such young clusters, the timescales for single-star evolution imply that the remnant population consists almost exclusively of neutron stars (NS) or black holes (BH). Consequently, any WD detection in these OCs would fundamentally challenge established WD formation theories. 

Our sample contains 16 extremely young OCs ($<25$ Myr)
We identified 46 WDs in these OCs, with 6, 13, and 27 satisfying the 5D, 3D-only, and 2D-only membership criteria, respectively. All WDs are newly discovered, with 33/46 potentially formed through binary evolution. The WD distributions in these OCs are shown in Figure Set 2 (see Appendix \ref{appendix B}), with their corresponding candidate properties summarized in Table \ref{tab:elmocwd}.

\textbf{Collinder 69:}
In this cluster, we identified three WD candidates belonging to the cluster membership (one satisfying the 2D membership criteria, one satisfying the 3D criteria, and one satisfying the 5D membership criteria). All these WDs might form via binary evolution. Collinder 69 is a compact OC located in the center of the star-forming region Lambda Orionis (LOSFR; \citealt{Neha2024A&A...689A.225N}). The WD candidate satisfying the 5D selection criteria ( \textit{Gaia} EDR3 3337904688163790080) discovered in this cluster warrants significant follow-up attention. Given the cluster's young age of approximately 19 Myr, this WD is highly unlikely to have formed through single-star evolution. This target most likely originated from binary evolution or is a WD binary system.



\textbf{HSC 2733:}
In this cluster, we identified three WD candidates that are probable cluster members (two satisfying the 2D membership criteria and one ( \textit{Gaia} EDR3 5999490866325281664) satisfying the 5D membership criteria). All these WDs meet the selection criteria for having undergone binary evolution. For the WD that satisfies the 5D membership criteria,  \textit{Gaia} XP spectrum classification suggests it is likely a DQ WD, with an estimated effective temperature of $T_{\mathrm{eff}} = 7,791$ K and a mass of 0.593 $M_\odot$ \citep{Vincent2024A&A...682A...5V}. Its relatively low temperature makes it an unlikely product of a binary merger \citep{Kawka2023MNRAS.520.6299K}. However, single-star evolution models cannot account for its formation within the cluster age, suggesting a binary origin. This object represents a compelling candidate for follow-up.


\textbf{UPK 422:}
In this cluster, we identified two WD candidates ( \textit{Gaia} EDR3 3018549920968071424,  \textit{Gaia} EDR3 3022162847457598080) that satisfy the 5D membership criteria.  Both targets have low probabilities of being WDs (based on \cite{2021MNRASGentileFusillo}) and lie outside the region covered by WD cooling models. They also exhibit large astrometric uncertainties. Based on their positions on the CMD, these objects may be WD--MS binary systems.


\textbf{Haffner 13:}
In this cluster, we identified 4 WD candidates belonging to the cluster(one satisfying the 2D membership criteria, two satisfying the 3D membership criteria, and one satisfying the 5D membership criteria). Among them, three WDs satisfy the selection criteria for having undergone binary evolution; two meet the 3D membership criteria and one meets the 5D membership criteria). The WD  \textit{Gaia} EDR3 5598427168110776832, which satisfies the 5D membership criteria, has a cooling age of 121 Myr. This significantly exceeds the age of its host cluster (20 Myr), suggesting a possible binary evolution origin. It is, therefore, an important target for follow-up observations. 

\textbf{IC 2391:}
In this cluster (a triaxial ellipsoidal cluster \citep{Pang2021ApJ...912..162P}), we identified three WD candidates belonging to the cluster (one satisfying the 2D membership criteria and two satisfying the 3D membership criteria). Among them, two WDs satisfy the binary evolution selection criteria (with one meeting the 2D membership criteria and one meeting the 3D membership criteria). The target ( \textit{Gaia} EDR3 5317454079808243456), which meets the 2D membership criteria, has been identified by \cite{Richer2021ApJ...912..165R} as a wide-search member of the cluster (see Table 4 of \cite{Richer2021ApJ...912..165R}).


\subsection{Young open clusters}
We classify OCs with ages between 25 Myr and 100 Myr as young OCs, as only massive stars within this age range have had sufficient time to evolve into WDs through single-star evolution (progenitor mass $\gtrsim 6.3~{M_\odot}$, yielding WD mass $\gtrsim 1.1~M_\odot$), making them exceptionally rare in these clusters. Theoretically, any WDs present in such young clusters formed through single-star evolution should be of high mass. However, owing to the low luminosity resulting from their early formation and extensive cooling, these high-mass WDs are difficult to detect observationally. Consequently, given these constraints on single-star evolution, binary interactions are expected to be a dominant formation channel in young OCs.


Our sample contains 37 young OCs (25--100 Myr).
We identified 136 WDs in these OCs, with 21, 31, and 84 satisfying the 5D, 3D-only, and 2D-only membership criteria, respectively. Among these 136 WDs, 95 WDs satisfy the binary evolution constraint (with 54 meeting the 2D membership criteria,  23 meeting the 3D membership criteria, and 18 meeting the 5D membership criteria). The expected number of WDs that may have evolved from single stars according to our calculations is 67. The number of WDs we find associated with clusters that satisfy single-star evolution is 41, some of which have inaccurate probability calculations owing to large astrometric errors. Even after accounting for these inaccurate members, the number of WDs we find with single star evolution is smaller than the expected value, suggesting that some WDs may have escaped from the cluster.  The WD distributions in these OCs are shown in Figure Set 3 (see Appendix \ref{appendix B}), with their corresponding candidate properties summarized in Table \ref{tab:yocwd}.

\textbf{Alessi 34:}
In this cluster, we identified 12 WD candidates that are probable cluster members (8 satisfying the 2D membership criteria, 3 satisfying the 3D membership criteria, and 1 satisfying the 5D membership criteria). All these WDs meet the selection criteria for having undergone binary evolution.  \textit{Gaia} EDR3 5493134460004973184 (a 5D member) is a candidate for an ELM WD \citep{Pelisoli2019MNRAS.488.2892P}. Its cooling age is 151 Myr, far exceeding the age of the cluster (35 Myr). 

\begin{figure}[h!]
\centering
\includegraphics[width=0.45\textwidth]{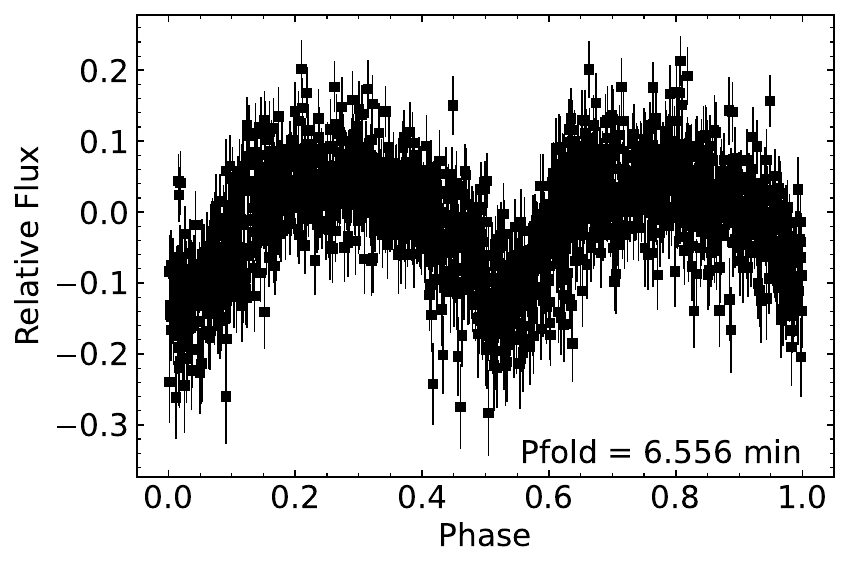}
\caption{RSG5-WD light curve. The period of RSG5-WD is 6.556 minutes. 
\label{figrsg5}}
\end{figure}

\textbf{RSG 5:}
In this cluster, we identified one WD candidate ( \textit{Gaia} EDR3 2082008971824158720) that satisfies the 5D membership criteria and might form via binary evolution. RSG 5 is a very young OC. The age of RSG 5 has been estimated by several studies, ranging from 22 to 57 million years \citep{Hunt2024A&A...686A..42H,Almeida2023MNRAS.525.2315A,Bouma2022AJ....164..215B}.  \textit{Gaia} EDR3 2082008971824158720 (the 5D member) was also classified as a cluster member by \cite{Bouma2022AJ....164..215B},\cite{ Hunt2023A&A...673A.114H} and \cite{Prisegen2023A&A...678A..20P} using clustering algorithm. 
By cross-matching with Zwicky Transient Facility (ZTF) photometric data and applying the Lomb--Scargle period search algorithm \citep{VanderPlas2018ApJS..236...16V}, we detected a variability period of 6.556 minutes. Figure \ref{figrsg5} shows the light curve of this WD. As shown by \cite{Yan2025ApJ...991L...7Y}, a binary merger provides a plausible formation channel for this star within the cluster's age.

\textbf{Platais 9:}
In this cluster, we identified 14 WD candidates belonging to the cluster (11 satisfying the 2D membership criteria, one satisfying the 3D membership criteria, and 2 satisfying the 5D membership criteria). Among them, 10 WDs satisfy the selection criteria for having undergone binary evolution; 7 meet the 2D membership criteria, one meets the 3D membership criteria and 2 meet the 5D membership criteria). The two WDs, \textit{Gaia} EDR3 5429378904609527808 (4.58 Gyr) and \textit{Gaia} EDR3 5620660732036506240 (1.35 Gyr), satisfy the 5D membership criteria and have cooling ages that differ by about 3 Gyr. Both vastly exceed the cluster's age. We note that \textit{Gaia} EDR3 5429378904609527808 was previously identified as a candidate member by \cite{miller2025whitedwarfinitialfinalmass} but was excluded from detailed follow-up in their study due to its significant age discrepancy with the cluster. Nevertheless, the anomalous ages of both objects make them compelling targets for further investigation to confirm their nature. 

\textbf{Melotte 20 (Alpha Persei):}
In this cluster, we identified seven WD candidates belonging to the cluster (two satisfying the 2D membership criteria, four satisfying the 3D membership criteria, and one satisfying the 5D membership criteria). All these WDs meet the selection criteria for having undergone binary evolution. The WD Gaia EDR3 246940317212992768, which satisfies the 5D selection criteria, has a cooling age of 3.81 Gyr. This far exceeds the age range of its host cluster (51Myr), strongly suggesting it originated from binary evolution. Consequently, it is a compelling target for follow-up observations. The remaining candidates include  \textit{Gaia} EDR3 244003693457188608 and  \textit{Gaia} EDR3 439597809786357248. 
These sources were previously identified as cluster members by \cite{Lodieu2019A&A...628A..66L} and \cite{Heyl2021arXiv211004296H}. 
Furthermore, \cite{Miller2022ApJ...926L..24M} classified them as runaway WDs that have escaped the cluster;
notably, \cite{Miller2022ApJ...926L..24M} obtained Gemini spectra for both targets, which provided spectroscopic support for their 
status as escaped members. Additionally,  \textit{Gaia} EDR3 435725089313589376 is considered a candidate 
WD of the cluster by \cite{Casewell2015MNRAS.451.4259C}. Based on their positions on the CMD, 
all three candidates appear to be in the early stages of cooling.

\textbf{UPK 562:}
In this cluster, we identified two WD candidates that satisfy the 2D membership criteria but not binary evolution criterion. One of them, \textit{Gaia} EDR3 5338712763507161728, was analyzed in the study by \cite{Raddi2016MNRAS.457.1988R}. They concluded that it is a background star based on a distance estimate of $1174\pm174$\,pc, which places it well behind NGC~3532 (distance $\sim 480$\,pc). Although \cite{Raddi2016MNRAS.457.1988R} noted that the object could be closer, even a significant shift in $\log g$ results in a distance of $\sim 800$\,pc, confirming that it is certainly not a member of NGC~3532. Regarding UPK~562 (distance $\sim 834$\,pc in \cite{Hunt2023A&A...673A.114H}), while it is reasonable to classify this object as a background field star at its nominal distance, a higher $\log g$ would place it at $\sim 800$\,pc, making it a plausible member of UPK~562. Based on its spectral features, \cite{Raddi2016MNRAS.457.1988R} found no distinct characteristics and classified it as a DC WD (Figure 3 of \cite{Raddi2016MNRAS.457.1988R}). However, their photometric fitting estimated its temperature to be around 29,500 K, which is atypical for a DC WD. We suggest this WD could also exhibit weak spectral features owing to the presence of a strong magnetic field \citep{Ferrario2015SSRv..191..111F,Raddi2016MNRAS.457.1988R}.

\subsection{Intermediate-age Open Clusters}
We classify open clusters with ages between 100 and 1000 Myr as intermediate-age clusters. WDs found in these systems may form through the evolution of single stars or as a result of binary evolution. Our sample includes 59 intermediate-age open clusters within this age range. 

We identified 226 WDs in these OCs. Among them, 66 satisfy the 5D membership criteria, 61 meet the 3D-only membership criteria, and 99 fulfill the 2D-only criteria. Of these 226 WDs, 113 WDs satisfy the binary evolution constraint, with 48 meeting the 2D membership criteria, 39 meeting the 3D membership criteria, and 26 meeting the 5D membership criteria. The estimate number of WDs expected to form via single-star evolution in these OCs is estimated to be between 579 and 729. This estimate is based on an upper mass limit of $8{-}12\,M_{\odot}$ for the most massive MS star that can evolve into a WD. Even after taking into account the number of WDs formed through binary evolution, the observed number of WDs in these clusters remains significantly lower than expected. This discrepancy suggests that the majority of WDs have already escaped from these clusters (under the assumption that our detected WD sample is complete).  The WD distributions in these OCs are shown in Figure Set 4 (see Appendix \ref{appendix B}), with their corresponding candidate properties summarized in Table \ref{tab:mocwd}.


\textbf{NGC 2422:}
In this cluster, we identified three WD candidates as cluster members. One candidate satisfies the 2D membership criteria and two satisfy the 3D membership criteria. Among these, only one of the 3D candidates ( \textit{Gaia} EDR3 3029894574568016512) meets the criterion for binary evolution.

The earliest search for WDs in NGC 2422 was carried out by \citet{Romanishin1980ApJ...235..992R} using UV observations. Subsequently, \citet{Koester1981A&A....99L...8K} identified a WD candidate ( \textit{Gaia} DR2 3030026344167186304) that might be associated with the cluster, although its nature could not be definitively determined. They suggested this source could be a massive WD associated with the cluster, a field WD situated behind the cluster, or a subdwarf O-type star. Later, \citet{Richer2019ApJ...880...75R} identified a massive WD with a helium-rich atmosphere and a strong magnetic field in the cluster, which likely forms a binary system with a late-type companion ( \textit{Gaia} DR2 3029912407273360512). The only WD recovered by \citet{Prisegen2021A&A...645A..13P} was the one previously reported by \citet{Richer2019ApJ...880...75R}; the other candidate from \citet{Koester1981A&A....99L...8K} was not included in the catalog of \citet{Gentile2019MNRAS.482.4570G}. Using  \textit{Gaia} DR2 astrometry, it was determined that the candidate from \citet{Koester1981A&A....99L...8K} is not a cluster member and lies in the foreground. In both our study and in the work by \citet{Prisegen2023A&A...678A..20P},  \textit{Gaia} DR2 3029912407273360512 was also not identified as a member of NGC 2422. 

\textbf{Stock 12:}
In this cluster, we identified two WD candidates that are likely cluster members. One candidate satisfies the 3D membership criteria, while the other meets the 5D membership criteria. Among these, only the 5D WD ( \textit{Gaia} EDR3 199246910423973) fulfills the selection criterion for binary evolution. Previous studies have searched for and analyzed WDs in Stock 12. It is worth noting that this source was also independently identified as a cluster member and included in the recent IFMR analysis by \citet{miller2025whitedwarfinitialfinalmass}. Both \citet{Prisegen2021A&A...645A..13P} and \citet{Prisegen2023A&A...678A..20P} reported the same WD in this cluster, corresponding to  \textit{Gaia} EDR3 1992469104239732096, which is the 5D member identified in our study. \citet{Richer2021ApJ...912..165R} also detected this WD in their search for massive WDs coinciding with OCs using the  \textit{Gaia} DR2 database. Notably, this source was also independently identified as a cluster member and included in the recent IFMR analysis by \cite{miller2025whitedwarfinitialfinalmass}.

\begin{figure*}[!ht]
\centering
\includegraphics[width=0.85\textwidth]{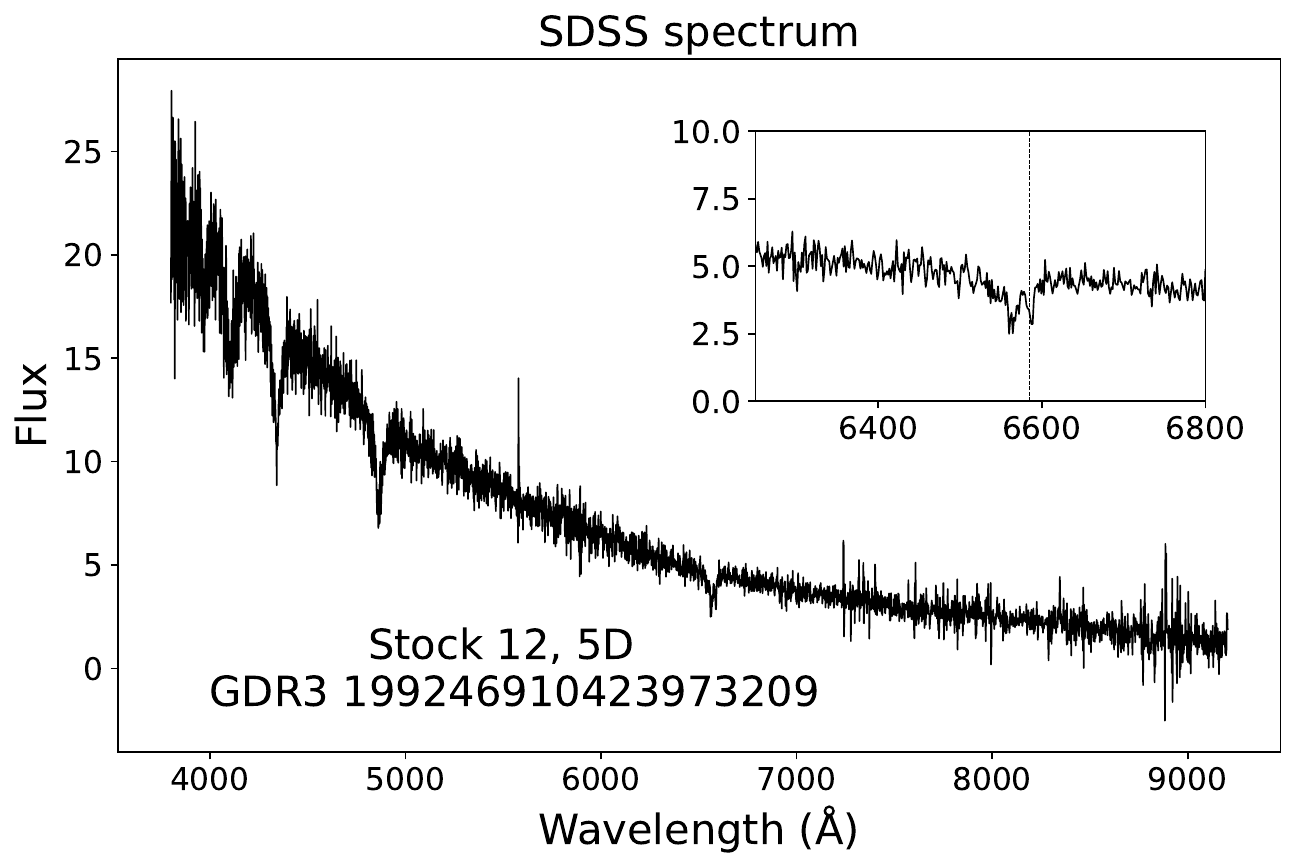}\\
\includegraphics[width=0.85\textwidth]{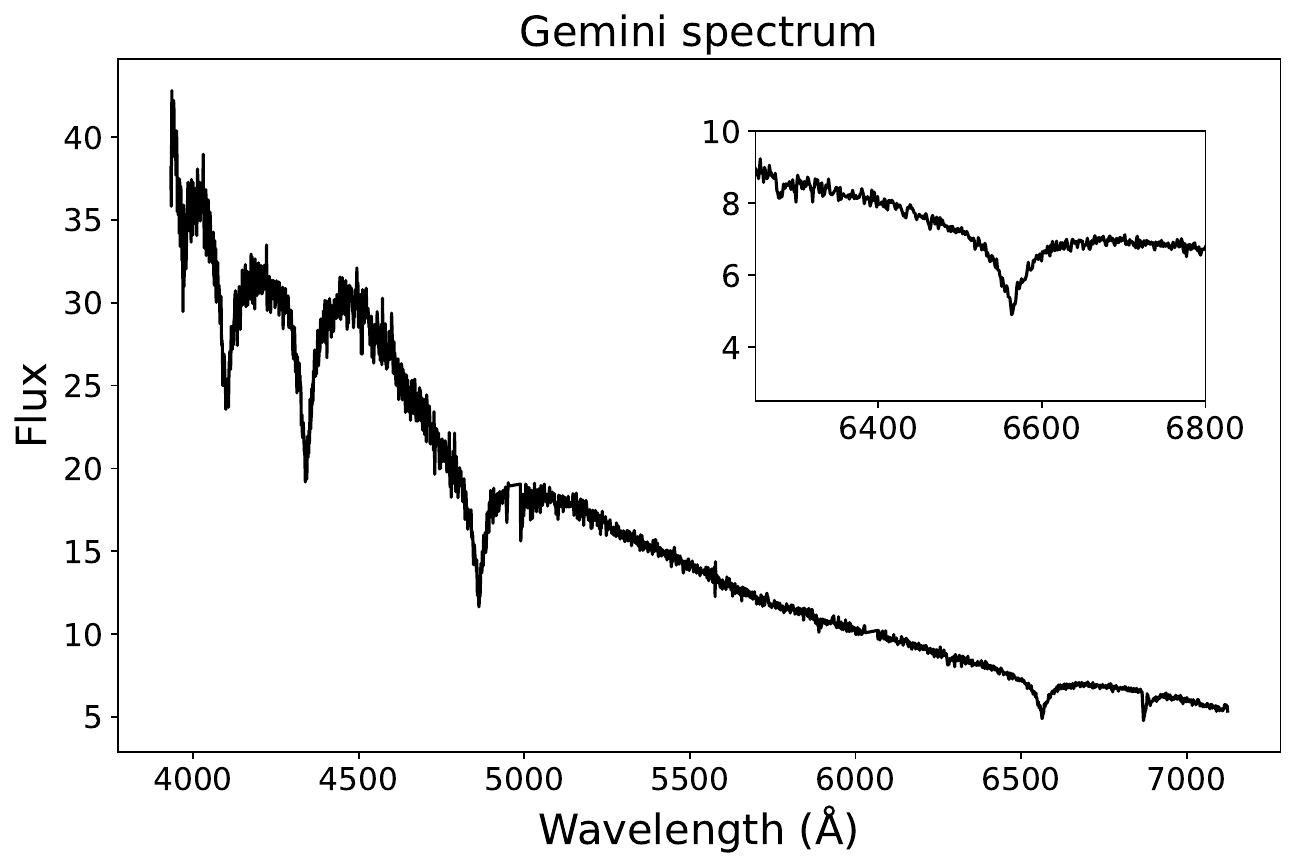}\\
\caption{The SDSS spectrum\citep{Kleinman2013ApJS..204....5K} (top panel) and the Gemini spectrum\citep{Richer2021ApJ...912..165R} (bottom panel) of  \textit{Gaia} EDR3 1992469104239732096. Each panel's inset shows the spectrum near the $\text{H}\alpha$ absorption line. In the top panel, the black dashed line indicates the position of the Zeeman split.
\label{Gaia1992469104239732096}}
\end{figure*}

 \textit{Gaia} EDR3 1992469104239732096 was previously identified as a suspected magnetic WD by \citet{Kleinman2013ApJS..204....5K} based on SDSS spectroscopy. In the SDSS spectrum, we observed Zeeman splitting near the $\text{H}\alpha$ absorption line. However, in the follow-up Gemini spectrum obtained by \citet{Richer2021ApJ...912..165R}, no significant magnetic features were detected.  Figure \ref{Gaia1992469104239732096} presents a comparison of the spectra from these two observations. Although the SDSS spectrum displays a doublet-like feature, it is absent in the Gemini spectrum, which has a significantly higher signal-to-noise ratio (S/N). Given that Zeeman splitting at $\text{H}\alpha$ typically manifests as a triplet and is not observed in the superior Gemini data, we attribute the feature in the SDSS spectrum to a noise artifact. Consequently, we conclude that this object is likely a non-magnetic WD.


%

\textbf{Theia 172:}
In this cluster, we identified 4 WD candidates as likely cluster members: two WDs satisfy the 2D membership criteria, one WD satisfies the 3D membership criteria, and one WD satisfies the 5D membership criteria. Only the two WDs selected by the 2D criteria satisfy the binary evolution criterion. The cluster is expected to host 2–4 WDs originating from single-star evolution. 
\citet{Prisegen2023A&A...678A..20P} identified a WD in this cluster, which matches the 5D member WD found in our study: Gaia EDR3 3114831641658036608. This source was also independently identified by \citet{miller2025whitedwarfinitialfinalmass}. Although its total age ($185$\,Myr) appears slightly older than our estimated cluster age ($121$\,Myr), \citet{miller2025whitedwarfinitialfinalmass} noted that this discrepancy is consistent with single-star evolution, as it requires only a $\sim 1\sigma$ shift in the cluster age to align the two. However, it is worth noting that \citet{miller2025whitedwarfinitialfinalmass} did not prioritize this object due to its substantial parallax uncertainty in the \textit{Gaia} data.

It is important to emphasize the significance of this 5D member. While it does not fully satisfy the strict binary evolution criterion ($P_\mathrm{bin} > 0.8$), it still has a substantial probability of having formed through the binary channel, with $P_\mathrm{bin} > 0.5$. We therefore suggest that this WD be given particular attention in future observational studies.

\textbf{NGC 1039:}
In this cluster, we identified two WD candidates as cluster members, both of which satisfy the 2D membership criteria. Notably, both WDs also meet the selection criterion for having undergone binary evolution. The first search for WDs in NGC 1039 (M34) was carried out by \citet{Rubin2008AJ....135.2163R}, who initially identified 17 potential WD members in this cluster; however, only 5 of these were found to be consistent with the cluster’s distance, while the remaining candidates were brighter and likely represent WD binaries or field stars.

In our study, two WDs previously identified by \citet{Rubin2008AJ....135.2163R} (\textit{Gaia} EDR3 337044088221827456 and \textit{Gaia} EDR3 337155723012394752, referred to as LAWDS~S2 and LAWDS~17, respectively) have been reported as cluster members and satisfy our 2D membership criteria. These two objects are also listed in \citet{Cummings2018ApJ...866...21C} and \citet{Richer2021ApJ...912..165R}, although they did not fulfill the more stringent selection criteria of \citet{Richer2021ApJ...912..165R}. This exclusion was primarily due to significant data limitations in \textit{Gaia} DR2: LAWDS~17 exhibited astrometric errors comparable to its proper motion and parallax values, while LAWDS~S2 lacked astrometric data entirely. In \textit{Gaia} DR3, although full astrometric solutions are now available, the uncertainties remain too high (particularly in $\mu_{\alpha}$) to fairly assess membership based solely on \textit{Gaia} parameters. However, it is worth noting that \citet{miller2025whitedwarfinitialfinalmass} included both objects as non-\textit{Gaia} sources and found them to be in good agreement with the IFMR, suggesting they are likely single-star evolutionary members. A definitive astrometric assessment will likely be possible with the release of \textit{Gaia} DR4 or DR5.



\textbf{Theia 274:}
In this cluster, we identified one WD candidate ( \textit{Gaia} EDR3 5529347562661865088) that satisfies the 5D membership criteria. This WD does not appear to have formed via binary evolution and can be considered a typical cluster member. Previous studies, including \citet{Richer2021ApJ...912..165R} and \citet{Caiazzo2020ApJ...901L..14C}, have associated this object with ASCC~47 (which may be the same cluster as Theia~274). Notably, spectra obtained by \citet{Caiazzo2020ApJ...901L..14C} revealed it to be a magnetic DA WD; however, its properties remain consistent with membership resulting from single-star evolution. Consequently, this object was included in the IFMR analysis of \cite{Richer2021ApJ...912..165R} and  \cite{miller2025whitedwarfinitialfinalmass}.

\begin{figure*}[!ht]
	\centering
	\includegraphics[width=1.0\textwidth]{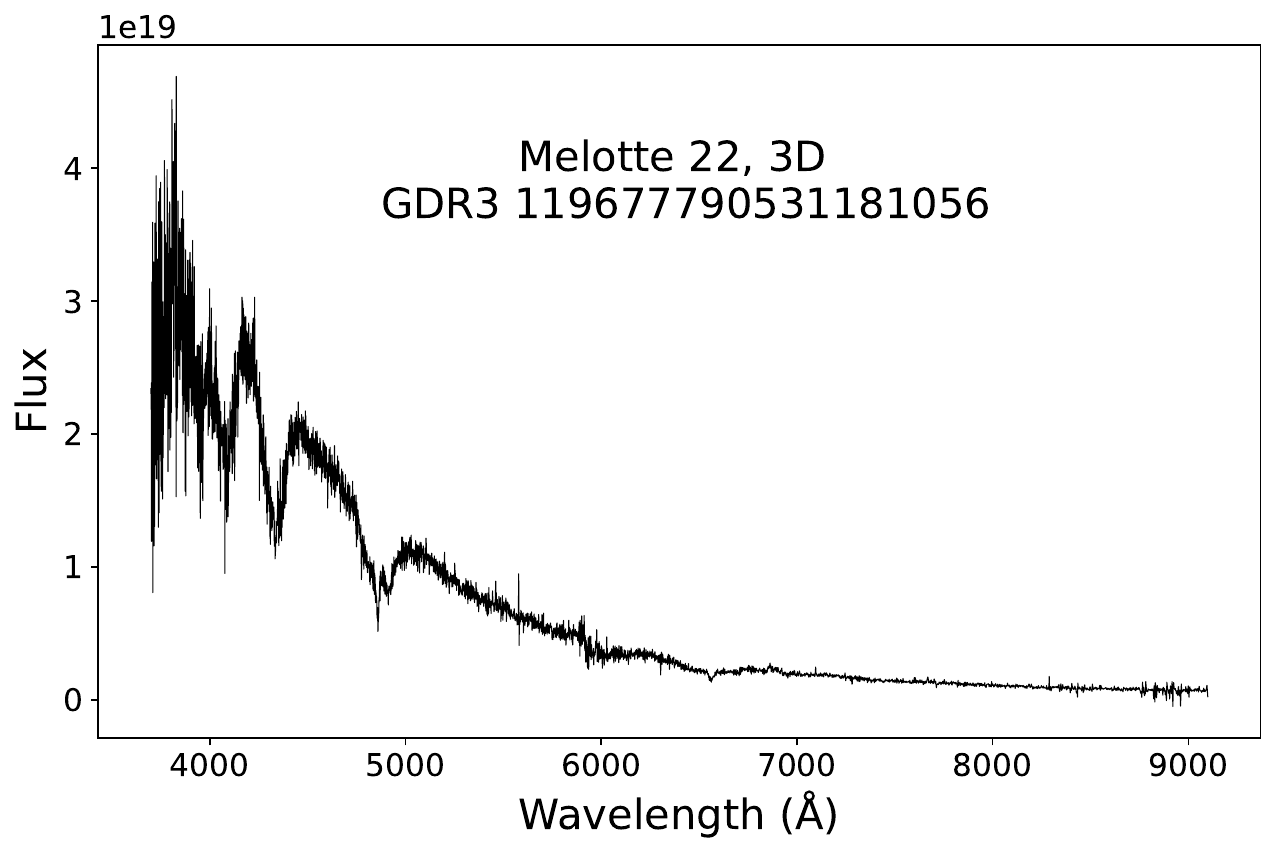}
	\caption{LAMOST spectrum of  \textit{Gaia} EDR3 119677790531181056 \citep{Guo2015MNRAS.454.2787G}.
		\label{Gaia119677790531181056}}
\end{figure*}

\textbf{Melotte 22:}
In this cluster, we identified 11 WD candidates that are probable cluster members (10 WDs satisfying the 3D membership criteria and one WD satisfying the 5D membership criteria). All these 3D WDs meet the selection criteria for having undergone binary evolution. Melotte 22 (Pleiades; M45) is a well-known and extensively studied OC. Its proximity  makes it visible to the naked eye. In previous studies, only one WD has been identified: GDR2 66697547870378368 \citep{Eggen1965ApJ...141...83E, Lodieu2019A&A...628A..66L, Prisegen2021A&A...645A..13P, Richer2021ApJ...912..165R,miller2025whitedwarfinitialfinalmass}. In our search, we also identified this target (a 5D member). The total age of this 5D WD member (147 Myr), calculated from single-star evolutionary models, is consistent with its host cluster's age of 150 Myr. The WD candidates found in Melotte 22 are basically in a sequence, suggesting they may have undergone the same evolutionary process, albeit at different times.  \textit{Gaia} EDR3 119677790531181056 (a 3D member) is identified as a magnetic WD based on its spectrum \citep{Guo2015MNRAS.454.2787G,Zhao2013AJ....145..169Z}. Figure \ref{Gaia119677790531181056} shows the LAMOST spectrum of this target, where the Zeeman splitting is clearly visible in the $\mathrm{H}\beta$ absorption line. 

\textbf{NGC 2516:}
We identified 7 WD candidates as possible cluster members: one satisfies the 2D membership criteria, one satisfies the 3D criteria, and five satisfy the 5D criteria. Among these, two WDs meet both the membership and binary evolution criteria: one 3D member ( \textit{Gaia} EDR3 5339470430053546240, newly identified) and one 5D member(\textit{Gaia} EDR3 5290719287073728128), both representing high-priority targets for further study. 

Our results are consistent with previous surveys \citep[e.g.,][]{Koester1996A&A...313..810K, Cummings2018ApJ...866...21C, Prisegen2021A&A...645A..13P, Prisegen2023A&A...678A..20P,miller2025whitedwarfinitialfinalmass}, and we independently recover all previously reported WD members in NGC 2516. Despite this, we find significantly fewer WDs than predicted by single-star evolution models, which expect at least 16 WDs in the cluster, suggesting that many cluster WDs have likely escaped or they remain undetected.


\textbf{Theia 517:}
In this cluster (Messier 39, NGC 7092), we identified 14 WD candidates as possible members: nine satisfy the 2D membership criteria, three satisfy the 3D criteria, and two satisfy the 5D criteria. Among these, 11 WDs also meet the binary evolution criterion, including eight 2D members, two 3D members, and one 5D member.

The 5D member satisfying the binary evolution criterion (\textsl{Gaia} EDR3 2170776080281869056) is of particular interest, as it was also independently reported as a cluster member by \citet{Prisegen2023A&A...678A..20P} and \citet{Miller2022ApJ...926L..24M}. Detailed spectral analysis by \citet{Caiazzo2020ApJ...901L..14C} further revealed evidence for a magnetic field, which they argued is consistent with single-star evolution. Adopting this conclusion, \citet{miller2025whitedwarfinitialfinalmass} subsequently incorporated this object into their study of the IFMR. However, our analysis suggests a binary evolutionary origin for this target, contrary to the single-star scenario proposed by \citet{Caiazzo2020ApJ...901L..14C} and adopted by \citet{miller2025whitedwarfinitialfinalmass}.

Notably, \textsl{Gaia} EDR3 1973737725372862464 (a 2D member with $P_\mathrm{bin}>0.8$) was identified as a DZ WD based on its \textsl{Gaia} XP spectrum \citep{Vincent2024A&A...682A...5V}.


\textbf{NGC 3532:}
We identified 22 WD candidates as possible cluster members: 12 satisfy the 2D membership criteria, three the 3D criteria, and 7 the 5D criteria. Of these, 5 WDs also meet the binary evolution criterion: three are 2D members, one is a 3D member, and one is a 5D member. Despite our expanded census, the total number of cluster WD candidates we identify remains well below the $\gtrsim$50 WDs predicted by single-star evolution, indicating that the majority of WDs have likely escaped from the cluster or they remain undetected.

Early surveys based on photographic plates identified several blue WD candidates in NGC 3532 \citep{Reimers1989A&A...218..118R, Koester1993A&A...275..479K}. Spectroscopic follow-up identified only a subset as \textit{bona fide} cluster WDs, leading to at most seven 
 members \citep{Dobbie2009MNRAS.395.2248D, Dobbie2012MNRAS.423.2815D, Raddi2016MNRAS.457.1988R}.  \textit{Gaia}-based studies by \citet{Gentile2019MNRAS.482.4570G} and \citet{Prisegen2021A&A...645A..13P}, with more stringent criteria and limited  \textit{Gaia} DR2 coverage, found only three WDs, some owing to incomplete sample cross-matching. The most recent and comprehensive census by \citet{Prisegen2023A&A...678A..20P} added five new  \textit{Gaia} EDR3 WDs and recovered several previous cluster members.

Our survey independently recovers the cluster member candidates reported by \citet{Prisegen2023A&A...678A..20P}. Regarding spectroscopic validation, \citet{miller2025whitedwarfinitialfinalmass} obtained Gemini spectra for four of these candidates that lacked prior literature data, confirming all four as DA WDs. Based on their analysis, one object (\textsl{Gaia} DR3 5340530320614001792) was ruled out as a single-star evolution member due to its very low mass and excessively old cooling age. The other three---comprising two members consistent with the cluster isochrone (\textsl{Gaia} DR3 5339402672685054720 and 5340288226187644160) and one questionable candidate (\textsl{Gaia} DR3 5340165355769599744)---were subsequently included in their study. Together with seven previously characterized members, this resulted in a total sample of ten WDs for their IFMR analysis.


\textbf{Stock 2}
In the Stock 2, we identified 27 WD candidates as possible members: 10 satisfy the 2D membership criteria, two the 3D criteria, and 15 the 5D criteria. Among these, 15 candidates also meet the binary evolution criterion, comprising four 2D, two 3D, and nine 5D members. Of the total 27 identified WDs, 13 were previously reported in earlier works.

Stock 2 has historically been understudied due to observational challenges, including its large angular extent and spatially variable reddening, which complicate membership determination \citep{Spagna2009MmSAI..80..129S}. Previous studies identified eight WD candidates based on \textsl{Gaia} DR2 data \citep{GaiaCollaboration2018A&A...616A...1G}, a census subsequently expanded to 16 candidates (including 10 new objects) by \citet{Prisegen2021A&A...645A..13P}, though two earlier candidates were not recovered. More recently, \citet{Prisegen2023A&A...678A..20P} refined the census, confirming seven probable WD members. Notably, \citet{Richer2021ApJ...912..165R} highlighted \textsl{Gaia} EDR3 506862078583709056 as a massive WD member of this cluster.

Spectroscopic follow-up was recently conducted by \citet{miller2025whitedwarfinitialfinalmass}, who obtained new spectra for nine WDs in this cluster using Gemini (seven targets) and Keck (two targets). Their analysis identified seven objects as non-magnetic DA WDs consistent with single-star evolution and cluster membership\footnote{The \textsl{Gaia} DR3 IDs are: 507054806657042944, 507119265523387136, 459270649787942784, 507128332197081344, 458778927573447168, 508276703371724928, and 507899399087944320.} Of the remaining two, one (\textsl{Gaia} DR3 507105143670906624) was identified as highly magnetic and was subsequently excluded from their IFMR analysis. The final target (\textsl{Gaia} DR3 511159317926025600) was classified as a metal-polluted DZA WD, likely indicating the accretion of debris from a disrupted planetary companion; however, spectral analysis revealed this object to be a non-member of the cluster(consistent with our own analysis). Consequently,
\citet{miller2025whitedwarfinitialfinalmass} included the seven identified non-magnetic DAs and the previously known massive WD from \citet{Richer2021ApJ...912..165R} in their IFMR determination.

Despite this expanded census, our estimates suggest that 29--36 WDs should have formed via single-star evolution in Stock 2, indicating a persistent deficit and the likely escape of WDs over time.

\textbf{NGC 2632:}
We identified 18 WD candidates in this cluster: two satisfy the 2D membership criteria, three the 3D criteria, and 13 the 5D criteria. Among these, 11 also fulfill the binary evolution criterion, including two 2D, three 3D, and six 5D members.

NGC 2632(also named M44 and Praesepehas) one of the best-studied WD populations of all open clusters. Previous comprehensive surveys by \citet{Salaris2019MNRAS.483.3098S} and \citet{Prisegen2021A&A...645A..13P} reported 12 WD members. Among the sample from \citet{Salaris2019MNRAS.483.3098S}, 11 objects possess pre-existing literature spectra and are thus identified members. The 12th candidate (\textsl{Gaia} DR3 662998983199228032), originally a new identification, has been recovered in subsequent works but currently lacks spectroscopic confirmation of its membership. Our study reproduces all 12 known members (each recovered as a 5D member in our analysis) and additionally identifies six new high-probability WD candidates. Among the newly found candidates, three 3D members display proper motion deviations, suggesting possible field contamination.

\begin{figure*}[!ht]
\centering
\includegraphics[width=0.9\textwidth]{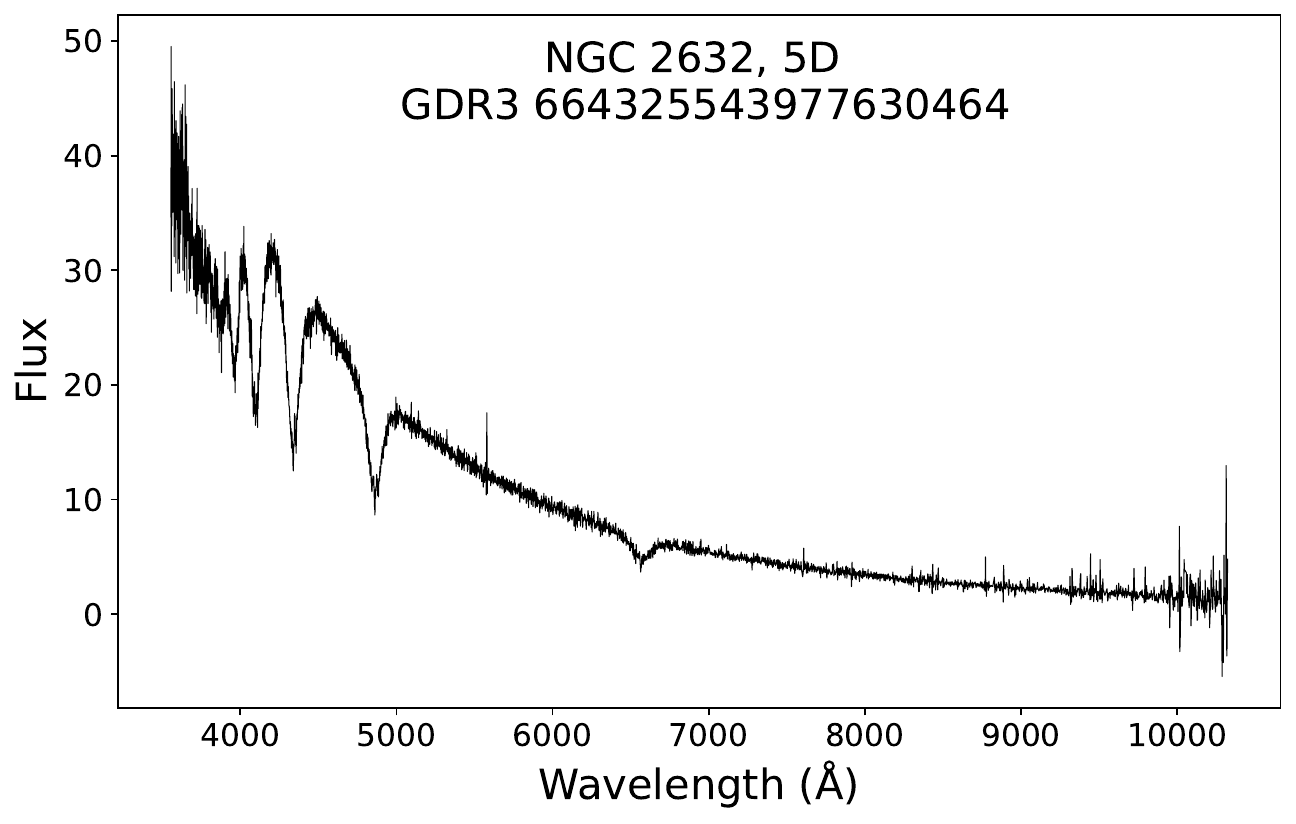}\\
\includegraphics[width=0.9\textwidth]{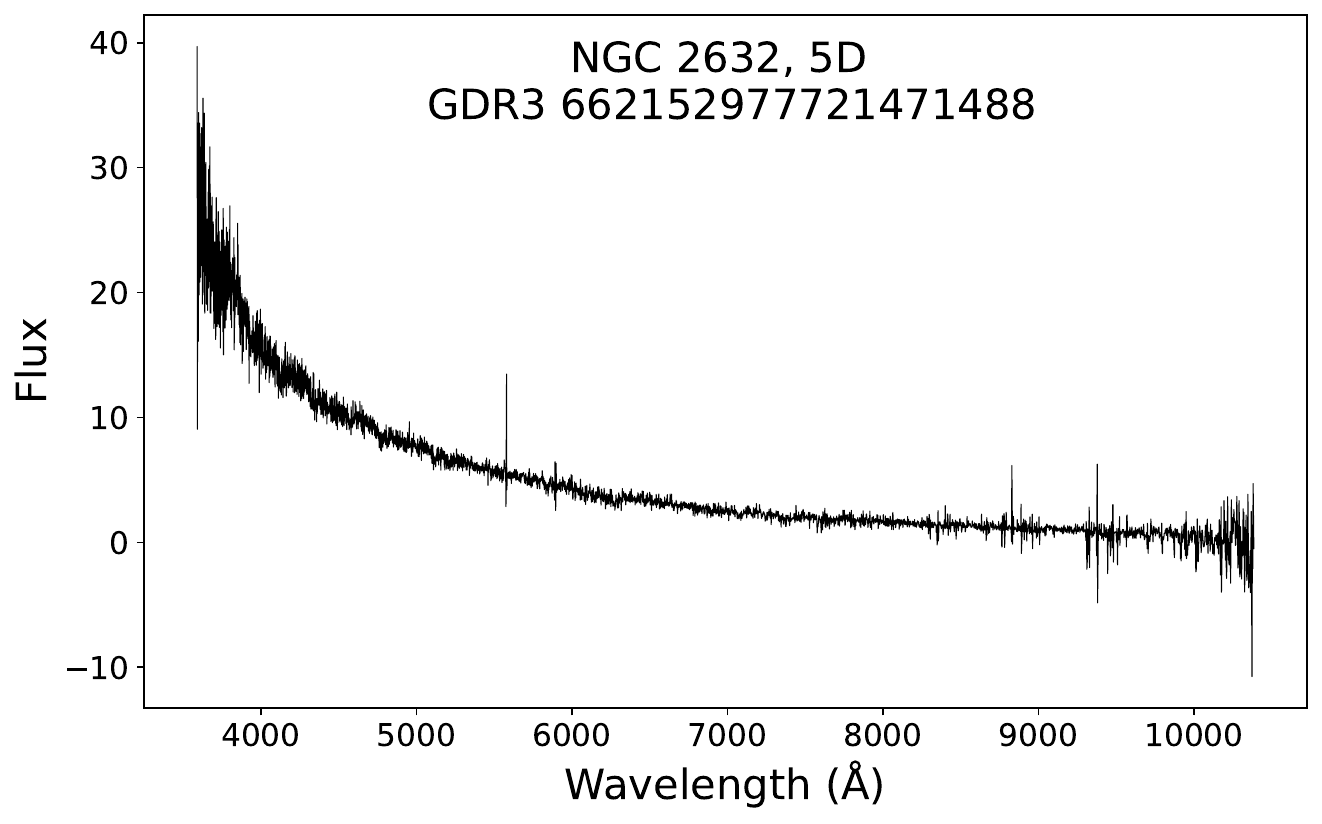}\\
\caption{SDSS spectra of  \textit{Gaia} EDR3 664325543977630464 (top panel) \citep{Kleinman2013ApJS..204....5K} and
 \textit{Gaia} EDR3 662152977721471488 (bottom panel)\citep{Vincent2024A&A...682A...5V}. 
\label{GaiaEDR664325543977630464}}
\end{figure*}

\textsl{Gaia} EDR3 664325543977630464, which meets our binary-evolution criterion, is a spectroscopically identified magnetic WD \citep{Kleinman2013ApJS..204....5K}. \citet{miller2025whitedwarfinitialfinalmass} included this target in their study, noting that its position on the IFMR is consistent with a single-star origin. However, given its magnetism and kinematics, a binary evolutionary history cannot be entirely ruled out.  \textit{Gaia} EDR3 662152977721471488 is classified as a DC WD based on SDSS, yet its measured temperature (19,597 K; \citealt{Vincent2024A&A...682A...5V}) is unusually high for this type—suggesting a strong magnetic field may have suppressed spectral features. SDSS spectra for both objects are shown in Figure~\ref{GaiaEDR664325543977630464}.

\begin{figure*}[!ht]
\centering 
\includegraphics[width=1.0\textwidth]{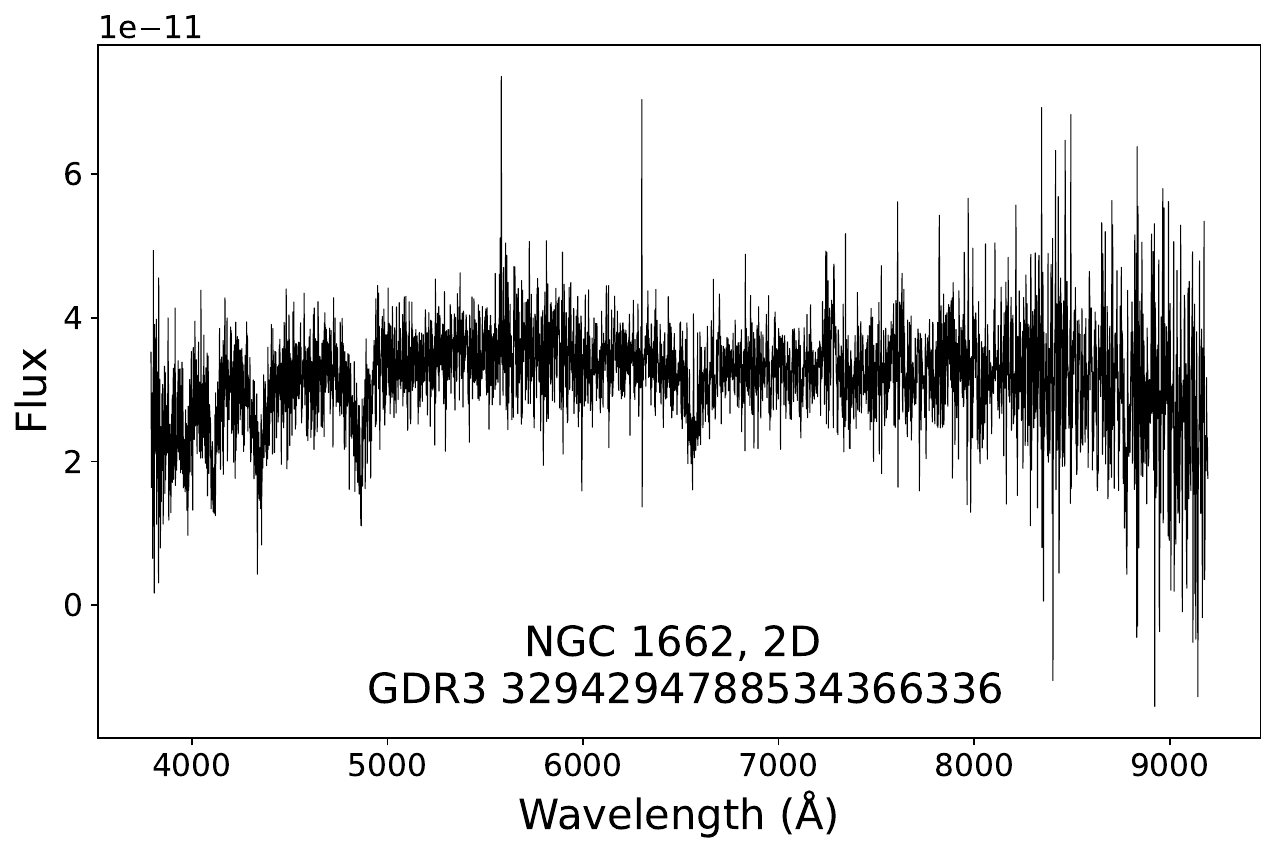}
\caption{SDSS spectrum of  \textit{Gaia} EDR3 3294294788534366336 \citep{Kleinman2013ApJS..204....5K}. 
\label{GaiaDR3294294788534366336}}
\end{figure*}

\textbf{NGC 1662:}
We identified one WD candidate in this cluster that satisfies the 2D membership criteria but does not meet the binary evolution criterion. Although it only meets the 2D membership selection, its stellar parameters are highly consistent with those of the cluster. Spectroscopic analysis of SDSS data \citep{Kleinman2013ApJS..204....5K} confirms this object as a magnetic DA WD, as illustrated in Figure~\ref{GaiaDR3294294788534366336}.

\textbf{Alessi 62:}
We identified one WD candidate in this cluster that satisfies the 5D membership criteria but does not meet the binary evolution condition. Two WDs have previously been reported as associated with Alessi 62. The first,  \textit{Gaia} DR2 4519349757791348480, was noted by \citet{Prisegen2021A&A...645A..13P} but with a low WD classification probability ($P_\mathrm{wd}=0.56$), and it is absent from the latest  \textit{Gaia} EDR3 WD catalog \citep{2021MNRASGentileFusillo}. 

The second object, \textsl{Gaia} EDR3 4519349757798439936, was reported as a member candidate based on EDR3 data by \citet{Prisegen2023A&A...678A..20P}. While also identified in \citet{miller2025whitedwarfinitialfinalmass}, did not prioritize this target for spectroscopic follow-up. This object appears to be an extremely young WD, and its derived properties are particularly sensitive to the assumed reddening. Consequently, while it is a likely single-star member under current assumptions, a small shift in reddening could allow for binary evolutionary scenarios.

\subsection{Old open clusters}

We classify OCs older than 1000 Myr as old OCs. At such ages, their most massive surviving stars (8–10~$M_{\odot}$) will have completed MS evolution and evolved into compact remnants, while intermediate-mass stars will have exhausted their nuclear fuel, shed their envelopes as planetary nebulae, and collapsed into WDs. The combined effects of stellar evolution and long-term dynamical relaxation selectively deplete high-mass members—either through transformation into remnants or dynamical ejection—resulting in a stellar population dominated by low-mass stars, a characteristic feature distinguishing old OCs from younger clusters.

Our sample contains five old OCs ($>1000$ Myr).
We identified a total of 31 WDs in these OCs: 18 satisfy the 5D membership criteria, 4 satisfy only the 3D criteria, and 9 satisfy only the 2D criteria. Among these, 17 WDs are newly identified, and 3 WDs satisfy the binary evolution criterion (with one meeting the 2D, one the 3D, and one the 5D criteria). Based on single-star evolution models, we estimate these OCs should collectively host 597–729 WDs, including 100–112 predicted in NGC 5822 and 233–244 in NGC 2682. Notably, for all old OCs in our sample, the predicted WD populations substantially exceed the observed numbers. This is largely expected, as the oldest WDs in these clusters have likely cooled below the detection limits.
The WD distributions in these OCs are shown in Figure Set 5 (see Appendix \ref{appendix B}), with their corresponding candidate properties summarized in Table \ref{tab:oocwd}.

\textbf{NGC 752:}
In this cluster, we identified 6 WD candidates consistent with the cluster: two satisfying the 2D membership criteria, three satisfying the 3D criteria, and one satisfying the 5D criteria. Only the 5D candidate ( \textit{Gaia} EDR3 342523646152426368) meets the binary evolution criterion. \cite{miller2025whitedwarfinitialfinalmass} also identified this 5D candidate.
\citet{Buckner2018RNAAS...2..151B} previously identified a potential WDMS binary member in this cluster based on  \textit{Gaia} DR2 proper motions and parallax, suggesting that, if confirmed, it would be the first known WD+MS system in NGC 752. In our study, we also recovered  \textit{Gaia} EDR3 343129953800001408. Although this source satisfies the initial 5D criteria, it was excluded from our final WD list due to its low WD probability ($P_\mathrm{wd}=0.015$). Additionally, ZTF photometric data show no significant periodic variability for this object.

\textbf{IC 4756:}
In this cluster, we identified three WD candidates consistent with cluster membership: one satisfying the 2D membership criteria and two satisfying the 5D criteria. Only the 2D candidate meets the binary evolution selection criterion. IC 4756 is a nearby and old OC whose WD population, although the cluster has been extensively studied, has only recently been explored in detail thanks to  \textit{Gaia} data. Notably, \textsl{Gaia} EDR3 4283928577215973120 satisfies the 5D membership criteria and was previously reported as a member candidate by \citet{Prisegen2021A&A...645A..13P} and \citet{Prisegen2023A&A...678A..20P}. \citet{miller2025whitedwarfinitialfinalmass} also identified this target but excluded it from spectroscopic follow-up given its expected low mass. Similarly, regarding \textsl{Gaia} DR3 4284010735661963648, \citet{miller2025whitedwarfinitialfinalmass} identified this object but excluded it from subsequent study due to high astrometric uncertainty.

\textbf{NGC 6991:}
In this cluster, we identified 5 WD candidates as probable members: three satisfying the 2D membership criteria, one satisfying the 3D criteria, and one satisfying the 5D criteria. Only the 3D WD meets the binary evolution selection criterion. \citet{Prisegen2021A&A...645A..13P,Prisegen2023A&A...678A..20P,miller2025whitedwarfinitialfinalmass} reported a WD candidate ( \textit{Gaia} EDR3 2166915179559503232) potentially associated with this cluster. Our analysis shows that this WD candidate satisfies the 2D membership criteria. The improved  \textit{Gaia} DR3 astrometry suggests it may indeed be associated with NGC 6991, contrary to its initial exclusion based on  \textit{Gaia} DR2 parallax measurements in \citet{Prisegen2021A&A...645A..13P}, a conclusion now consistent with \citet{Prisegen2023A&A...678A..20P}.For the remaining 5D WD member, although it does not strictly meet the binary evolution criterion, it maintains a relatively high probability of a binary origin ($P_\mathrm{bin}>0.5$) and was also identified by \citet{miller2025whitedwarfinitialfinalmass}. Additionally, we note that \textsl{Gaia} DR3 2163824456685399168, which satisfies the 2D selection criteria, is also included in the \citet{miller2025whitedwarfinitialfinalmass} catalog.

%
%

\textbf{NGC 2682:}
In this cluster, we identified three WD candidates as cluster members, all satisfying the 5D membership criteria. None of these WD members meet the binary evolution selection criterion. The study of WDs in NGC 2682 (M67) was pioneered by \citet{Bellini2010A&A...513A..50B}, who performed the first proper motion-based identification of cluster-associated WDs. \citet{Sindhu2018MNRAS.481..226S} subsequently expanded the sample by detecting additional members through UV observations. In our investigation, we identified three WDs in this cluster that satisfy the 5D membership criterion. 

These candidates were also examined by \citet{miller2025whitedwarfinitialfinalmass}, who provided detailed characterizations for their exclusion from the IFMR analysis. Specifically, \textsl{Gaia} DR3 604898490980773888 was spectroscopically confirmed as a DB WD. \textsl{Gaia} DR3 604721293514179712 exhibits an extremely low mass, indicative of a binary evolutionary origin. Finally, \textsl{Gaia} DR3 604917698073661952 is considered a photometric non-member in the literature.

\section{Discussion}\label{sec5}

\subsection{White Dwarf Natal Kicks}\label{sce6}
The number of observed WDs in OCs is significantly lower than predicted by theoretical models \citep{Richer2021ApJ...912..165R}. Our single-star evolutionary models, assuming progenitor masses in the range of 8--12~$M_{\odot}$, predict a total WD population of 1027--1254, which agrees with independent estimates for similar clusters \citep{Richer2021ApJ...912..165R}. This discrepancy between observations and predictions is most likely attributable to natal kick mechanisms during WD formation. For single stars, asymmetric mass loss during the asymptotic giant branch (AGB) phase can impart a natal kick to the remnant \citep{Fellhauer2003ApJ...595L..53F, Kalirai2008ApJ...676..594K}. In binary systems, \citet{Sandquist1998ApJ...500..909S} demonstrated that asymmetric CE ejection can impart characteristic kicks of 3--8~km~s$^{-1}$ to post-CE products. Such evolution processes in binary systems can thus efficiently expel WDs from clusters.

In line with this scenario, \citet{Grondin2024arXiv240704775G} identified WD+MS systems whose proper motions are consistent with cluster membership, yet which display significant spatial offsets—just as expected from natal kicks. Their analysis demonstrates that typical kick velocities result in much greater positional displacements than changes in proper motion. This suggests that many WDs identified via 2D selection criteria (showing proper-motion consistency but spatial offsets) may be genuine cluster members that have experienced natal kicks during their formation. Given that the majority of WDs subjected to natal kicks are expected to retain parallaxes consistent with that of the parent cluster, we examined candidates showing offsets in coordinate space. We required their parallax values to align with the cluster's mean parallax within a $3\sigma$ interval, accounting for measurement errors. Out of the 123 candidates that initially satisfied the 2D cluster membership selection, 63 met this parallax criterion. Consequently, these 63 objects are considered probable escaped WDs that have retained kinematic association with the cluster. Potentially, in the case of older and nearby clusters, natal kicks may lead to parallax inconsistencies due to spatial dispersion along the line of sight.

\begin{figure*}[!ht]
\centering
\includegraphics[width=0.85\textwidth]{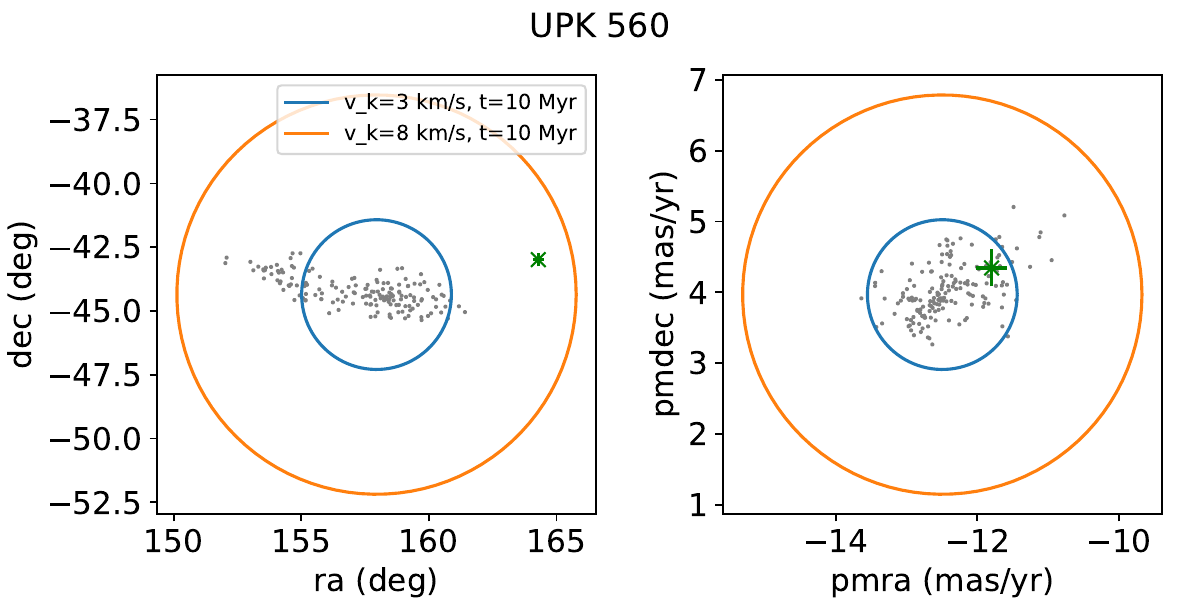}\\
\includegraphics[width=0.85\textwidth]{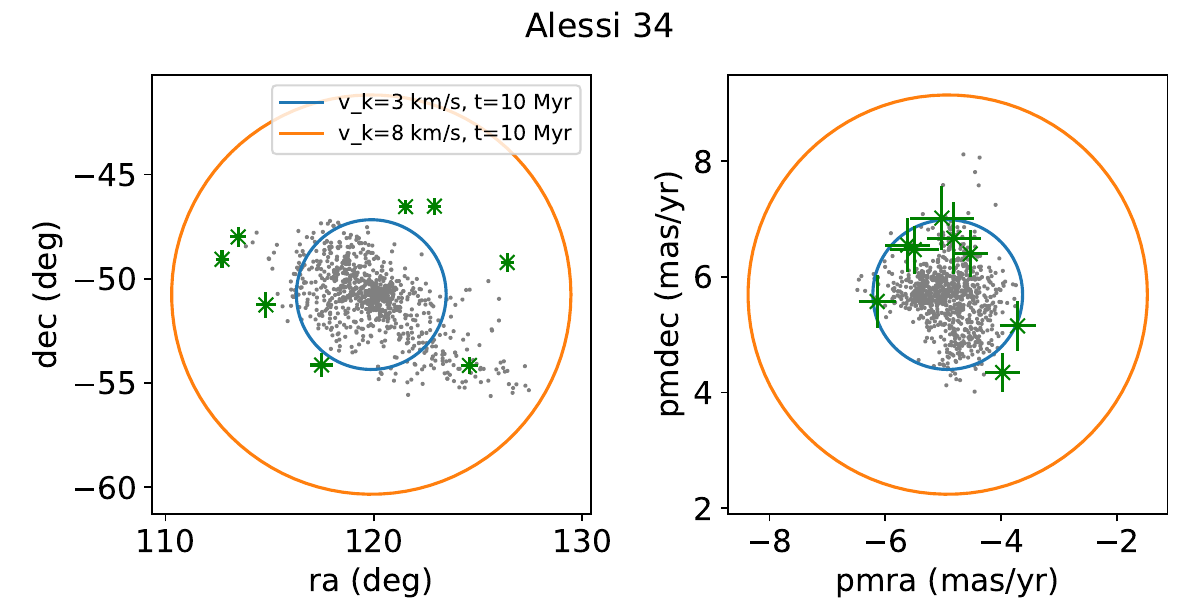}\\
\caption{The proposed effect of a CE-induced natal velocity kick on the spatial positions and kinematics of UPK 560 and Alessi 34. Left: Spatial distributions of cluster members, with each circle indicating the predicted displacement from the cluster center for different kick magnitudes and travel times. The green points represent the WD member candidates. Right: Similar to the left panel, but showing the corresponding proper motion distributions.
\label{fig:kick}}
\end{figure*}

Following the methodology of \citet{Grondin2024arXiv240704775G}, we assess this effect by comparing spatial displacements and changes in proper motion for typical kick velocities (3--8~km~s$^{-1}$). As illustrated in Figure~\ref{fig:kick} for two example systems (UPK~560 and Alessi~34), spatial offsets are significantly larger than the corresponding changes in proper motion at a given kick velocity. The observed kinematic properties of these spatially offset WD candidates support the hypothesis that they may have originated from CE-induced natal kicks. We will conduct a comprehensive study of binary-evolved WDs that may have escaped from clusters in future work.

Observations reveal that the number of WDs detected in OCs is often lower than predicted by theoretical models \citep{Weidemann1977A&A....59..411W,Richer1998ApJ...504L..91R,Richer2021ApJ...912..165R}. While dynamical escape certainly contributes to the loss of WDs, observational biases may also contribute to the observed deficit. A critical factor to consider is the completeness of the \textit{Gaia} sample. Although most of the WDs detectable by \textit{Gaia} can be found out to $\sim 500$\,pc, the 100-pc sample remains the most nearly volume-complete representation of the local population \citep{JimenezEsteban2023MNRAS.518.5106J}. Furthermore, the spectroscopic census is almost complete within only 40\,pc \citep{Tremblay2020MNRAS.497..130T, McCleery2020MNRAS.499.1890M, OBrien2023MNRAS.518.3055O, OBrien2024MNRAS.527.8687O}. At larger distances, many objects still lack spectroscopic confirmation, despite recent progress from automatic classification methodologies \citep{Vincent2024A&A...682A...5V, GarciaZamora2025A&A...699A...3G}. Given that the OCs analyzed in this work are located at significantly larger distances (up to $\sim 1.1$\,kpc), the incompleteness of the sample for faint, evolved  WDs is expected to be substantial, which likely contributes to the observed deficit. Similarly, binary interactions can accelerate the formation of WDs by stripping the stellar envelope, causing the remnant to enter the cooling sequence prematurely. This premature entry results in an extended cooling duration, potentially leading to derived cooling ages that are significantly larger than those predicted by single-star evolution scenarios. Furthermore, in addition to binary interactions, internal physical processes such as core crystallization significantly impact the cooling timescales. The release of latent heat and gravitational energy upon crystallization leads to a cooling delay, manifesting as the so-called Q branch on the CMD \citep{Tremblay2019Natur.565..202T,Cheng2019ApJ...886..100C}.  This effect is particularly relevant for massive WDs, including those formed via mergers. Such cooling delays can make these objects appear younger and more luminous than standard cooling models predict, further complicating the reconciliation of WD cooling ages with the cluster age

\section{Summary}\label{sec6}
In this paper, we search for WDs from the \cite{2021MNRASGentileFusillo} catalog that may have originated from binary evolution channels in OCs selected from the \cite{Hunt2024A&A...686A..42H} catalog. Our main results are summarized as follows:

\begin{enumerate}
  \item We identified 439 WD candidates associated with 117 OCs. Of these, 244 are likely to have originated from binary evolution. Within this subset, 49 are high-confidence 5D cluster members, including 32 that are newly identified.

  \item We cross-matched the WD candidates discovered in the OCs with spectroscopic data from LAMOST, SDSS, and Gemini, as well as time-domain photometry from ZTF. Among these candidates, several exhibit noteworthy properties, such as exceptionally strong surface magnetic fields, rapid optical variability on timescales of minutes. These features provide strong evidence for the important role of binary evolution channels in shaping the observed WD population in OCs.
  
  \item Approximately 50\% of the WD binary candidates have proper motions consistent with their host clusters and satisfy the 2D membership criteria. The observed spatial offsets may be explained by natal kicks received during the CE ejection or AGB phases. We find that moderate natal kicks of 3--8~km\,s$^{-1}$ can account for the population of WDs showing significant spatial offsets but consistent proper motions. 

\end{enumerate}

\section{Acknowledgments}
\begin{acknowledgments}
This project is supported by the National Key R\&D Program of China (2020YFC2201400) and the National Natural Science Foundation of China (NSFC grant Nos. 12033013). HY thanks the CAST Youth Science and Technology Talent Cultivation Program for Doctoral Students. HG acknowledges supports by the NSFC (Nos. 12288102, 12525304) and the 
Strategic Priority Research Program of the Chinese Academy of Sciences (No. 
XDB1160201). We would like to acknowledge Yang Xiang and Yifan Wang for their helpful discussions.
\end{acknowledgments}

\appendix\label{SectionA}
\renewcommand{\thefigure}{A.\arabic{figure}} 
\setcounter{figure}{0}
\section{Appendix A: WDs in OCs identified based on Gaia DR2 WD catalog} \label{Appendix A}

We used the same method to search for WDs in OCs, utilizing the Gaia DR2 WD sample selected by \cite{Gentile2019MNRAS.482.4570G}. We identified 44 WDs in 29 OCs. The WDs identified in clusters CWNU 519, HSC 2403, Melotte 20, RSG 4, Stock 2, Theia 96, and UPK 51 using the Gaia DR2 WD sample \citep{Gentile2019MNRAS.482.4570G} are all recovered in the Gaia DR3 WD catalog \citep{2021MNRASGentileFusillo}. We present WDs identified using only the Gaia DR2 WD sample in the Figure Set 1.

For the open clusters Collinder 69, NGC 6716, Theia 3397, UBC 11, UBC 32, Theia 553, Collinder 394, Teutsch 35, NGC 6025, Theia 181, Alessi-Teutsch 12, Alessi 40, Theia 643, Ferrero 1, and UPK 567, the WDs discovered within them are present only in the Gaia DR2 WD sample and have been excluded from the Gaia DR3 WD sample. We find that, based on Gaia DR3 photometric data, these candidate WDs clearly deviate from the WD sequence, suggesting that most of them are likely not genuine single WDs. Some of them may be binary systems containing a WD and a main-sequence star (WD+MS binaries).

For HSC 2636, we identified 10 samples that meet the 2D selection criteria based on the Gaia DR3 WD sample. Using the Gaia DR2 WD sample, we found only one sample that satisfies the 2D selection criteria. This target is not present in the Gaia DR3 WD sample, but it falls within the WD region on the HR diagram. For BH164, we found a WD that meets the 3D selection criteria. However, its proper motion is very inconsistent with that of the cluster, and it only appears in the Gaia DR2 WD sample. This target is likely contaminated by a field star. For Theia 401, we did not find any WDs belonging to the cluster in the Gaia DR3 WD sample. Based on the Gaia DR2 WD sample, we identified a WD that meets the 2D selection criteria. 

\figsetstart
\label{figgaiadr2}
\figsetnum{1}
\figsettitle{WDs in OCs identified based on Gaia DR2 WD catalog}

\figsetend


\begin{table}[h]
	\centering
	\caption{WD Member Candidates in OCs from \textit{Gaia} DR2 (This table is available in its entirety in machine-readable form.)}
	\label{tab:dr2wd}
	\begin{tabular}{cccccccc}
		\hline \hline
		Gaia DR2 ID & label & oc\_name & cooling age & mass & log Teff & log g & probability \\
		& & & (Gyr) & ($M_{\odot}$) & (K) & (cm\,s$^{-2}$) & \\ \hline
		4092374792459782912 & 2D & Alessi\_40 & & & & & 0.000 \\
		4519349757791348480 & 5D & Alessi\_62 & 0.007 & 0.246 & 4.177 & 7.017 & 1.000 \\
		5915271818988123776 & 2D & Alessi-Teutsch\_12 & & & & & 0.000 \\
		5848959241930068736 & 3D & BH\_164 & 1.732 & 0.236 & 3.897 & 7.109 & 0.512 \\
		4086034630451912832 & 3D & Collinder\_394 & & & & & 0.005 \\
		$\vdots$ & $\vdots$ & $\vdots$ & $\vdots$ & $\vdots$ & $\vdots$ & $\vdots$ & $\vdots$ \\
		4093247014448389760 & 3D & UBC\_32 & & & & & 0.030 \\
		6903997028802066816 & 2D & UPK\_51 & 0.025 & 0.243 & 4.121 & 7.041 & 0.878 \\
		5289776760792342144 & 2D & UPK\_545 & 0.182 & 0.461 & 4.129 & 7.712 & 1.000 \\
		5296990592159143552 & 5D & UPK\_545 & & & & & 0.000 \\
		5228492216046774656 & 2D & UPK\_567 & & & & & 0.038 \\ \hline
	\end{tabular}
	\begin{flushleft}
		\small
		\textbf{Notes:} \textit{Gaia} DR2 ID: Unique source identifier; oc\_name: Name of the host OC; cooling age: Cooling age in Gyr; mass: Mass in $M_\odot$; log Teff: $\log_{10} [T_{\mathrm{eff}}/\text{K}]$; log g: $\log_{10} [g/(\text{cm}\,\text{s}^{-2})]$; probability: Calculated probability of formation through binary evolution.
	\end{flushleft}
\end{table}

\section{Appendix B}\label{appendix B}
\renewcommand{\thefigure}{B.\arabic{figure}} 
\setcounter{figure}{0}

\begin{table}[h]
	\centering
	\caption{Parameters of the  OCs (This table is available in its entirety in machine-readable form.)}
	\label{tab:oc_para}
	\begin{tabular}{cccccccc}
		\hline \hline
		oc\_name & age & RAdeg & DEdeg & pmRA & pmDE & Plx & expect\_wd\_number \\
		& (Myr) & (deg) & (deg) & (mas\,yr$^{-1}$) & (mas\,yr$^{-1}$) & (mas) & \\ \hline
		OC\_0185 & 9 & 332.741 & 63.396 & -1.301 & -2.486 & 1.077 & 0 \\
		Pozzo\_1 & 9 & 122.375 & -47.354 & -6.364 & 9.437 & 2.847 & 0 \\
		UPK\_41 & 11 & 279.462 & 0.461 & 2.657 & -8.269 & 2.092 & 0 \\
		Collinder\_69 & 13 & 83.814 & 9.888 & 1.215 & -2.047 & 2.504 & 0 \\
		HSC\_2636 & 13 & 204.889 & -44.482 & -25.972 & -19.412 & 7.472 & 0 \\
		$\vdots$ & $\vdots$ & $\vdots$ & $\vdots$ & $\vdots$ & $\vdots$ & $\vdots$ & $\vdots$ \\
		NGC\_752 & 1174 & 29.160 & 37.793 & 9.772 & -11.824 & 2.269 & 46 \\
		IC\_4756 & 1288 & 279.627 & 5.437 & 1.284 & -4.968 & 2.108 & 54 \\
		NGC\_6991 & 1548 & 313.649 & 47.377 & 5.604 & 8.436 & 1.767 & 45 \\
		Ruprecht\_147 & 3019 & 289.141 & -16.269 & -0.949 & -26.672 & 3.275 & 68 \\
		NGC\_2682 & 4265 & 132.850 & 11.817 & -10.965 & -2.906 & 1.150 & 244 \\ \hline
	\end{tabular}
	\begin{flushleft}
		\small
		\textbf{Notes:} oc\_name: Cluster name; age: Cluster age in Myr; RAdeg/DEdeg: Right Ascension and Declination in degrees; pmRA/pmDE: Proper motions in mas yr$^{-1}$; Plx: Parallax in mas; expect\_wd\_number\: Expected number of WDs in the cluster from single star evolution.
	\end{flushleft}
\end{table}

\figsetstart 
\label{fig: eyoc}
\figsetnum{2}
\figsettitle{WD Member Candidates in Extremely Young OCs}

\figsetend


\begin{table}[h]
	\centering
	
	\caption{WD Member Candidates in Extremely Young OCs (This table is available in its entirety in machine-readable form.)}
	\label{tab:elmocwd}
	\begin{tabular}{cccccccccc}
		\hline
		\hline
		Gaia DR3 ID & Pwd & label & oc\_name & cooling age & total age & mass & log Teff & log g & probability \\
		& & & & (Gyr) & (Gyr) & ($M_{\odot}$) & (K) & (cm\,s$^{-2}$) & \\ \hline
		3337904688163790080 & 0.994 & 5 & Collinder\_69 & 0.267 & 0.343 & 1.066 & 4.324 & 8.724 & 0.811 \\
		3334282312745679360 & 0.997 & 3 & Collinder\_69 & 0.241 & 0.366 & 0.974 & 4.296 & 8.568 & 0.966 \\
		3339243785951466496 & 0.996 & 2 & Collinder\_69 & 0.256 & 1.038 & 0.687 & 4.174 & 8.125 & 1.000 \\
		5598408201533040512 & 0.916 & 3 & Haffner\_13 & 0.290 & 254.792 & 0.339 & 4.016 & 7.446 & 0.923 \\
		5599665630519975424 & 0.999 & 3 & Haffner\_13 & 0.161 & 0.241 & 1.056 & 4.391 & 8.700 & 0.894 \\
		$\vdots$ & $\vdots$ & $\vdots$ & $\vdots$ & $\vdots$ & $\vdots$ & $\vdots$ & $\vdots$ & $\vdots$ & $\vdots$ \\
		3113259443108680576 & 0.982 & 2 & ZHBJZ\_1 & 0.751 & 1.303 & 0.731 & 4.021 & 8.210 & 0.999 \\
		3107179727923969024 & 0.999 & 2 & ZHBJZ\_1 & 0.110 & 0.147 & 1.214 & 4.538 & 9.017 & 0.727 \\
		5514619883578482304 & 0.977 & 3 & Pozzo\_1 & 0.798 & 2.228 & 0.623 & 3.970 & 8.040 & 1.000 \\
		2208661059085452160 & 0.999 & 3 & ASCC\_127 & 0.051 & 0.109 & 1.125 & 4.565 & 8.815 & 0.810 \\
		2207383533949456000 & 0.874 & 5 & ASCC\_127 & 0.568 & 4.979 & 0.547 & 3.999 & 7.908 & 0.990 \\ \hline
	\end{tabular}
	\begin{flushleft}
		\small
		\textbf{Notes:} \textit{Gaia} DR3 ID: Unique source identifier; Pwd: Probability of being a WD form \cite{2021MNRASGentileFusillo}; oc\_name: Name of the host OC; cooling/total age: Cooling and total age in Gyr; mass: Mass in $M_\odot$; log Teff: $\log_{10} [T_{\mathrm{eff}}/\text{K}]$; log g: $\log_{10} [g/(\text{cm}\,\text{s}^{-2})]$; probability: Calculated probability of formation through binary evolution.
	\end{flushleft}
\end{table}

\figsetstart
\label{fig: yoc}
\figsetnum{3}
\figsettitle{WD Member Candidates in Young OCs}

\figsetend

\begin{table}[h]
	\centering
	\caption{WD Member Candidates in Young OCs (This table is available in its entirety in machine-readable form.)}
	\label{tab:yocwd}
	\begin{tabular}{cccccccccc}
		\hline \hline
		Gaia DR3 ID & Pwd & label & oc\_name & cooling age & total age & mass & log Teff & log g & probability \\
		& & & & (Gyr) & (Gyr) & ($M_{\odot}$) & (K) & (cm\,s$^{-2}$) & \\ \hline
		4085127842611392512 & 0.676 & 3D & NGC\_6716 & 0.018 & 22.286 & 0.493 & 4.372 & 7.720 & 0.344 \\
		5515291719545721472 & 0.996 & 2D & Alessi\_34 & 0.466 & 0.777 & 0.836 & 4.139 & 8.363 & 0.998 \\
		5319832392174785792 & 0.999 & 3D & Alessi\_34 & 0.022 & 1.502 & 0.619 & 4.386 & 7.971 & 0.999 \\
		5319706777263666560 & 0.984 & 2D & Alessi\_34 & 0.338 & 0.795 & 0.759 & 4.159 & 8.241 & 0.988 \\
		5519394100510917504 & 0.999 & 2D & Alessi\_34 & 0.043 & 0.418 & 0.797 & 4.410 & 8.271 & 0.968 \\
		$\vdots$ & $\vdots$ & $\vdots$ & $\vdots$ & $\vdots$ & $\vdots$ & $\vdots$ & $\vdots$ & $\vdots$ & $\vdots$ \\
		3334282312745679360 & 0.997 & 2D & CWNU\_522 & 0.241 & 0.366 & 0.974 & 4.296 & 8.568 & 0.960 \\
		3241096155376300672 & 0.995 & 2D & CWNU\_522 & 0.007 & 0.157 & 0.945 & 4.584 & 8.485 & 0.825 \\
		3289249679430424448 & 0.990 & 2D & CWNU\_522 & 0.136 & 4.130 & 0.549 & 4.199 & 7.882 & 1.000 \\
		4518909124201481472 & 0.988 & 2D & UBC\_26 & 0.005 & 0.762 & 0.691 & 4.562 & 8.058 & 0.857 \\
		4108486005110686976 & 0.601 & 2D & ASCC\_87 & 0.329 & 446.982 & 0.343 & 4.435 & 7.195 & 0.525 \\ \hline
	\end{tabular}
	\begin{flushleft}
		\small
		\textbf{Notes:} Same as Table \ref{tab:elmocwd}.
	\end{flushleft}
\end{table}
\figsetstart
\label{fig: moc}
\figsetnum{4}
\figsettitle{WD Member Candidates in Intermediate-age OCs}

\figsetend


\begin{table}[h]
	\centering
	\caption{WD Member Candidates in Intermediate-age OCs (This table is available in its entirety in machine-readable form.)}
	\label{tab:mocwd}
	\begin{tabular}{cccccccccc}
		\hline \hline
		Gaia DR3 ID & Pwd & label & oc\_name & cooling age & total age & mass & log Teff & log g & probability \\
		& & & & (Gyr) & (Gyr) & ($M_{\odot}$) & (K) & (cm\,s$^{-2}$) & \\ \hline
		1976240931087864064 & 0.996 & 2D & Theia\_517 & 0.589 & 0.677 & 1.037 & 4.189 & 8.681 & 1.000 \\
		1977852059224547200 & 0.999 & 2D & Theia\_517 & 0.180 & 0.498 & 0.832 & 4.278 & 8.344 & 0.969 \\
		1977762449023773056 & 0.989 & 2D & Theia\_517 & 0.660 & 0.983 & 0.829 & 4.081 & 8.356 & 0.990 \\
		1980813937387143040 & 0.995 & 5D & Theia\_517 & 0.764 & 1.077 & 0.835 & 4.060 & 8.367 & 1.000 \\
		1973737725372862464 & 0.996 & 2D & Theia\_517 & 0.623 & 0.971 & 0.813 & 4.083 & 8.330 & 1.000 \\
		$\vdots$ & $\vdots$ & $\vdots$ & $\vdots$ & $\vdots$ & $\vdots$ & $\vdots$ & $\vdots$ & $\vdots$ & $\vdots$ \\
		3029894574568016512 & 0.980 & 3D & NGC\_2422 & 0.473 & 0.814 & 0.817 & 4.129 & 8.333 & 0.931 \\
		3030209584655720832 & 0.999 & 3D & NGC\_2422 & 0.054 & 0.091 & 1.216 & 4.625 & 9.015 & 0.451 \\
		5331558310019845888 & 0.944 & 2D & Pismis\_4 &  &  &  &  &  & 0.000 \\
		5329959310885368192 & 0.978 & 3D & Pismis\_4 & 0.109 & 208.143 & 0.358 & 4.143 & 7.461 & 0.938 \\
		5915848478476555776 & 0.684 & 3D & Alessi-Teutsch\_12 & 0.351 & 514.261 & 0.275 & 3.951 & 7.266 & 0.712 \\ \hline
	\end{tabular}
	\begin{flushleft}
		\small
		\textbf{Notes:} Same as Table \ref{tab:elmocwd}.
	\end{flushleft}
\end{table}

\figsetstart
\label{fig: ooc}
\figsetnum{5}
\figsettitle{WD Member Candidates in Older OCs}

\figsetgrpstart
\figsetgrpnum{5.1}
\figsetgrptitle{IC 4756}
\figsetplot{picture/ooc/IC_4756.pdf}
\figsetgrpnote{Same as Figure 2, but for }
\figsetgrpend

\figsetgrpstart
\figsetgrpnum{5.2}
\figsetgrptitle{NGC 2682}
\figsetplot{picture/ooc/NGC_2682.pdf}
\figsetgrpnote{Same as Figure 2, but for }
\figsetgrpend

\figsetgrpstart
\figsetgrpnum{5.3}
\figsetgrptitle{NGC 6991}
\figsetplot{picture/ooc/NGC_6991.pdf}
\figsetgrpnote{Same as Figure 2, but for }
\figsetgrpend

\figsetgrpstart
\figsetgrpnum{5.4}
\figsetgrptitle{NGC 752}
\figsetplot{picture/ooc/NGC_752.pdf}
\figsetgrpnote{Same as Figure 2, but for }
\figsetgrpend

\figsetgrpstart
\figsetgrpnum{5.5}
\figsetgrptitle{Ruprecht 147}
\figsetplot{picture/ooc/Ruprecht_147.pdf}
\figsetgrpnote{Same as Figure 2, but for }
\figsetgrpend

\figsetend


\begin{table}[h]
	\centering
	\caption{WD Member Candidates in Older OCs (This table is available in its entirety in machine-readable form.)}
	\label{tab:oocwd}
	\begin{tabular}{cccccccccc}
		\hline \hline
	     Gaia DR3 ID & Pwd & label & oc\_name & cooling age & total age & mass & log Teff & log g & probability \\
	& & & & (Gyr) & (Gyr) & ($M_{\odot}$) & (K) & (cm\,s$^{-2}$) & \\ \hline
		4184719200659984640 & 0.997 & 2D & Ruprecht\_147 & 0.110 & 1.799 & 0.601 & 4.243 & 7.969 & 0.361 \\
		4185582626535414912 & 0.957 & 2D & Ruprecht\_147 & 0.999 & 1.512 & 0.741 & 3.979 & 8.228 & 0.343 \\
		4183847562828165248 & 0.980 & 5D & Ruprecht\_147 & 0.349 & 3.290 & 0.572 & 4.084 & 7.940 & 0.527 \\
		4183978061110910592 & 0.970 & 5D & Ruprecht\_147 & 0.299 & 10.520 & 0.518 & 4.087 & 7.840 & 0.641 \\
		4183926006112672768 & 0.983 & 5D & Ruprecht\_147 & 0.429 & 1.401 & 0.665 & 4.087 & 8.099 & 0.290 \\
		$\vdots$ & $\vdots$ & $\vdots$ & $\vdots$ & $\vdots$ & $\vdots$ & $\vdots$ & $\vdots$ & $\vdots$ & $\vdots$ \\
		342523646152426368 & 0.993 & 5D & NGC\_752 & 0.208 & 63.632 & 0.460 & 4.112 & 7.714 & 0.821 \\
		342493302207091072 & 0.999 & 3D & NGC\_752 & 0.244 & 0.304 & 1.117 & 4.365 & 8.816 & 0.079 \\
		4283755889495255296 & 0.752 & 2D & IC\_4756 & 0.367 & 276.330 & 0.332 & 3.979 & 7.436 & 0.911 \\
		4283928577215973120 & 0.991 & 5D & IC\_4756 & 0.001 & 0.685 & 0.699 & 4.777 & 7.989 & 0.296 \\
		4284010735661963648 & 0.997 & 5D & IC\_4756 & 0.075 & 0.177 & 1.010 & 4.462 & 8.613 & 0.060 \\ \hline
	\end{tabular}
	\begin{flushleft}
		\small
		\textbf{Notes:} Same as Table \ref{tab:elmocwd}.
	\end{flushleft}
\end{table}

\vspace{5mm}

\bibliography{sample631}{}
\bibliographystyle{aasjournal}



\end{document}